\documentstyle[12pt]{article}
\begin{document}

\begin{titlepage}
\begin{flushright}
FERMILAB-Pub-94/382-T\\
SACLAY-SPHT-T94/157\\
DTP/94/102\\
\end{flushright}
\vspace{1cm}
\begin{center}
{\Large\bf The Inclusive Two Jet Triply Differential
Cross Section}\\
\vspace{1cm}
{\large
W.~T.~Giele}\\
\vspace{0.5cm}
{\it
Fermi National Accelerator Laboratory, P.~O.~Box 500,\\
Batavia, IL 60510, U.S.A.} \\
\vspace{1cm}
{\large
E.~W.~N.~ Glover}\\
\vspace{0.5cm}
{\it
Physics Department, University of Durham,\\ Durham DH1~3LE, England} \\
\vspace{0.5cm}
and \\
\vspace{0.5cm}
{\large
David~A.~Kosower}\\
\vspace{0.5cm}
{\it
Service de Physique Th\'eorique, Centre d'Etudes de Saclay,\\
F-91191 Gif-sur-Yvette cedex, France}\\
\vspace{0.5cm}
{\large November 7 1994}
\vspace{0.5cm}
\end{center}
\begin{abstract}
We study the inclusive two jet triply differential cross section
$d^3\sigma/dE_Td\eta_1d\eta_2$ at Fermilab energies.  Different
$\eta_1$ and $\eta_2$ pseudorapidity
regions are directly related to both the parton
level matrix elements and the parton densities at leading order.  
We present the
next-to-leading order [${\cal O}(\alpha_s^3)$] corrections 
and show that the shape of the distribution at fixed transverse energy
$E_T$ is a
particularly powerful tool for constraining the parton distributions
at small to moderate $x$ values. 
We investigate the renormalisation/factorisation scale uncertainty present 
in the normalisation and shape of the distribution
at next-to-leading order.
We discuss specific slices of the distribution,
the same-side/opposite side ratio and the signed pseudorapidity distribution,
in detail and compare them with preliminary experimental data.

\end{abstract}

\end{titlepage}

\section{Introduction}

Dijet production in hadron collisions occurs when two partons from the
incident hadrons undergo a hard pointlike interaction and scatter at
relatively large angles.  The two-jet cross section depends on both
the non-perturbative probability of finding a particular parton inside
the parent hadron and the dynamics of the hard scattering.  By
examining kinematic regions where the parton densities are well known,
we can probe the pointlike strong-interaction cross section.  One
example of this is the shape of the angular distribution of the jets
in their centre of mass frame.  Recent data provide clear evidence
that
a running coupling constant, as given by QCD, is needed to describe
the data \cite{Weerts} and that the
next-to-leading order QCD predictions
\cite{EKS2jet} are in good agreement with the data.
An alternative approach is to use the theoretical description of
the hard scattering to extract the distribution of
partons in the proton from the data.
This is particularly interesting since gluon scattering 
plays a very important role in two jet production,
and it may be possible to probe the gluon density in a more
direct way than is possible in deeply inelastic scattering
or in Drell-Yan processes.

The inclusive two-jet cross section can be described in terms of
variables most suited to the geometry of the detector; the transverse
energy of the leading jet, $E_T = E_{T1}$, and the pseudorapidities of
the two leading jets, $\eta_1$ and $\eta_2$.  Recently, the D0
collaboration has presented a preliminary measurement of
$d^3\sigma/dE_Td\eta_1 d\eta_2$
\cite{Weerts} as a function of $\eta_1$ and $\eta_2$ at fixed $E_T$.
This seemingly complicated three-dimensional quantity contains all
the information available from two jet events.  In particular, at
leading order, $\eta_1$ and $\eta_2$ are directly related to the
parton momentum fractions $x_1$, $x_2$,
\begin{eqnarray}
\label{xseqn}
x_{1} &=& \frac{E_T}{\sqrt{s}} \left ( \exp(\eta_1) + \exp(\eta_2 )
\right ),\nonumber\\
x_{2} &=& \frac{E_T}{\sqrt{s}} \left ( \exp(-\eta_1) + \exp(-\eta_2 )
\right ),\nonumber\\
\end{eqnarray}
so that a measurement of the triply differential cross section
$d^3\sigma/dE_Td\eta_1 d\eta_2$ at fixed $E_T$ corresponds to a
measurement of $d^2\sigma/dx_1dx_2$.  Although the overall
normalisation of cross sections is uncertain in perturbative QCD, one
might hope that the shape of this distribution is well-predicted and
that it can be used to discriminate between different parton
densities.  Inclusion of the next-to-leading order corrections
enhances the reliability of the calculation for both
shape and normalisation. 

Beyond leading order, however, Eq.~\ref{xseqn} is no longer satisfied and
the parton momentum fractions are only approximately determined by the
transverse energies and pseudorapidities of the jets.  A three-dimensional
plot may obscure some of the desired physics, so both the CDF \cite{CDFssos}
and D0 \cite{Weerts,D0sign} collaborations have focused on particular slices
of the general distribution.  The CDF collaboration has examined the ratio of
cross sections for same-side events ($\eta_1 \sim \eta_2$) to opposite-side
events ($\eta_1 \sim -\eta_2$) for different $E_T$ bins.  This reduces the
normalization uncertainty and enhances the small $x$ region, $x \sim
4E_T^2/s$.  At small transverse energies, this distribution can reliably
discriminate between singular ($xg(x) \sim x^{-0.5}$ at small $x$) and
non-singular $xg(x) \sim x^0$ behavior of the gluon
distribution~\cite{GGKssos}.  The D0 collaboration has taken slices in
$\eta_1$ which contain information over the whole $x$ range, $4E_T^2/s < x <
1$, but are more sensitive to the overall normalisation.  In this paper, we
study the shape of the distribution over the whole $\eta_1$--$\eta_2$
plane. First, we examine the triply-differential distribution at lowest
order.  We discuss how the available phase space grows as the number of final
state partons increases and relate different $\eta_1$ and $\eta_2$ regions to
both the parton level matrix elements and the parton densities (section~2).
The full next-to-leading order triply-differential cross section is presented
in section~3. We show that it is sensitive to the parton density functions
and indicate how the shape depends on the renormalisation
(and factorisation)
scale.  Our results are applied to the CDF same-side/opposite-side and D0
signed distributions in sections~4 and 5.  Finally, our main findings are
summarised in section~6.

\section{The ${\cal O}(\alpha_s^2)$ triply differential two jet
cross section}

The lowest order cross section is given by,
\begin{equation}
\frac{d^3\sigma}{ dE_Td\eta_1d\eta_2} = \frac{1}{8\pi}
 \sum_{ij}  x_1f_i(x_1,\mu_F)~x_2f_j(x_2,\mu_F)~
\frac{\alpha^2_s(\mu_R)}{E_T^3}
\frac{|{\cal M}_{ij}(\eta^*)|^2}{\cosh^4\eta^*},
\end{equation}
where $f_i(x,\mu_F)$ ($i=g,q,\bar q$) represents the density of parton
$i$ in the proton at factorisation scale $\mu_F$ and $|{\cal M}_{ij}|^2$
is the lowest order squared matrix element for $ij \to 2$~partons
summed and averaged over initial and final state spins and colours.
The strong couping constant $\alpha_s$ is evaluated at the 
renormalisation scale, $\mu_R$.
The parton level cross section is insensitive to Lorentz boosts and
$|{\cal M}_{ij}|^2$ therefore depends only on the parton pseudorapidity in
the parton-parton rest frame, $\eta^* = (\eta_1-\eta_2)/2$. To
understand how the cross section is distributed over the
$\eta_1-\eta_2$ plane, we recall the `single effective subprocess
approximation' \cite{sesa}.  In this approximation all parton-parton
scattering cross sections are taken to be equal but are weighted by
colour arguments.  Thus the gluon-gluon, quark-gluon and quark-quark
subprocesses are in the ratio $1 : \frac{4}{9} :
\left(\frac{4}{9}\right ) ^2$
so that,
\begin{equation}
\frac{d^3\sigma}{dE_Td\eta_1d\eta_2 } \sim   
\frac{1}{8\pi}~x_1F(x_1,\mu)~x_2F(x_2,\mu) 
~\frac{\alpha^2_s(\mu)}{E_T^3}
\frac{|{\cal M}_{gg}(\eta^*)|^2}{\cosh^4\eta^*},
\end{equation}
where $F(x,\mu)$ is the `single effective parton density',
\begin{equation}
F(x,\mu) = g(x,\mu) 
+ \frac{4}{9} \sum_q \left (q(x,\mu) + \bar q(x,\mu) \right ).
\end{equation}
A rough indication of how the physical cross section depends on
$\eta_1$ and $\eta_2$ can be obtained by studying the parton-parton
luminosity, $ x_1F(x_1,\mu)~x_2F(x_2,\mu)$, and the squared matrix
elements, $|{\cal M}_{gg}|^2/\cosh^4\eta^*$, for gluon-gluon
scattering separately.

\begin{figure}\vspace{8cm}
\includegraphics{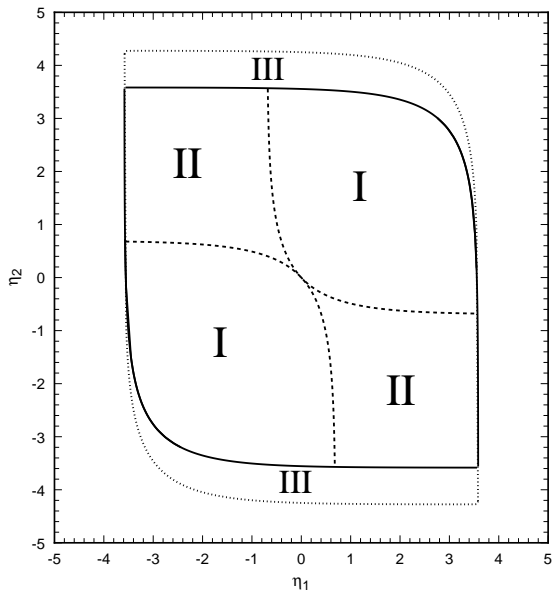}
\caption[]{The phase space boundary in the $\eta_1-\eta_2$ plane
at leading-order (solid) and next-to-leading order (dotted) for $E_T =
50$~GeV and $\sqrt{s} = 1800$~GeV.  The dashed line separates the
`small' $x$ -- `large' $x$ and `large' $x$ -- `large' $x$ regions.  In
region I, either $x_1$ or $x_2$ is less than $x_T$, while in region
II, both $x_1$ and $x_2$ are bigger than $x_T$.  Region III is only
permitted at next-to-leading order.}
\end{figure}

\def\etaboost{\eta_{\rm boost}}
However, to orient ourselves in the $\eta_1-\eta_2$ plane, we first
focus on the allowed phase space in terms of $\eta_1$ and $\eta_2$.  At lowest
order, the jet pseudorapidities are directly related to the parton
fractions via Eq.~1.  Since the momentum fraction cannot exceed unity,
we find,
\begin{equation}
-\log\left(\frac{2-x_T \exp(-\eta_1)}{x_T}\right ) < \eta_2 < 
\log\left( \frac{2-x_T\exp(\eta_1)}{x_T}\right),
\end{equation}
and,
\begin{equation}
|\eta_1| < \cosh^{-1} \left(\frac{1}{x_T}\right),
\end{equation}
where $x_T = 2E_T/\sqrt{s}$ and $x_T^2 < x_1 x_2 < 1$.  This boundary
is shown in Fig.~1 for $E_T = 50$~GeV and $\sqrt{s} = 1800$~GeV.  For
the opposite-side cross section, $\eta_1 \sim -\eta_2$, the parton
fractions are roughly equal so that in the top left and bottom right
corners of the allowed phase space, $x_1 \sim x_2 \to 1$.  On the other
hand, in the bottom left and top right corners, corresponding to
same-side events with $\eta_1 \sim \eta_2$, the parton fractions are
maximally different, $x_1 \to x_T^2$, $x_2 \to 1$ and vice versa.  The
dashed boundary separating region I and II makes a nominal division of
the phase space according to whether both parton fractions are `large'
or one parton fraction is `small'.  In region II, $x_1$ and $x_2 >
x_T$ while in region I, either $x_1 < x_T$ or $x_2 < x_T$.  The
corresponding axes for the pseudorapidities of the two jet system in the
laboratory, $\etaboost = (\eta_1+\eta_2)/2$, and of the jet in
the jet-jet center of mass frame, $\eta^* = (\eta_1-\eta_2)/2$, are
related by a rotation of $45^\circ$.

\begin{figure}\vspace{8cm}
\includegraphics{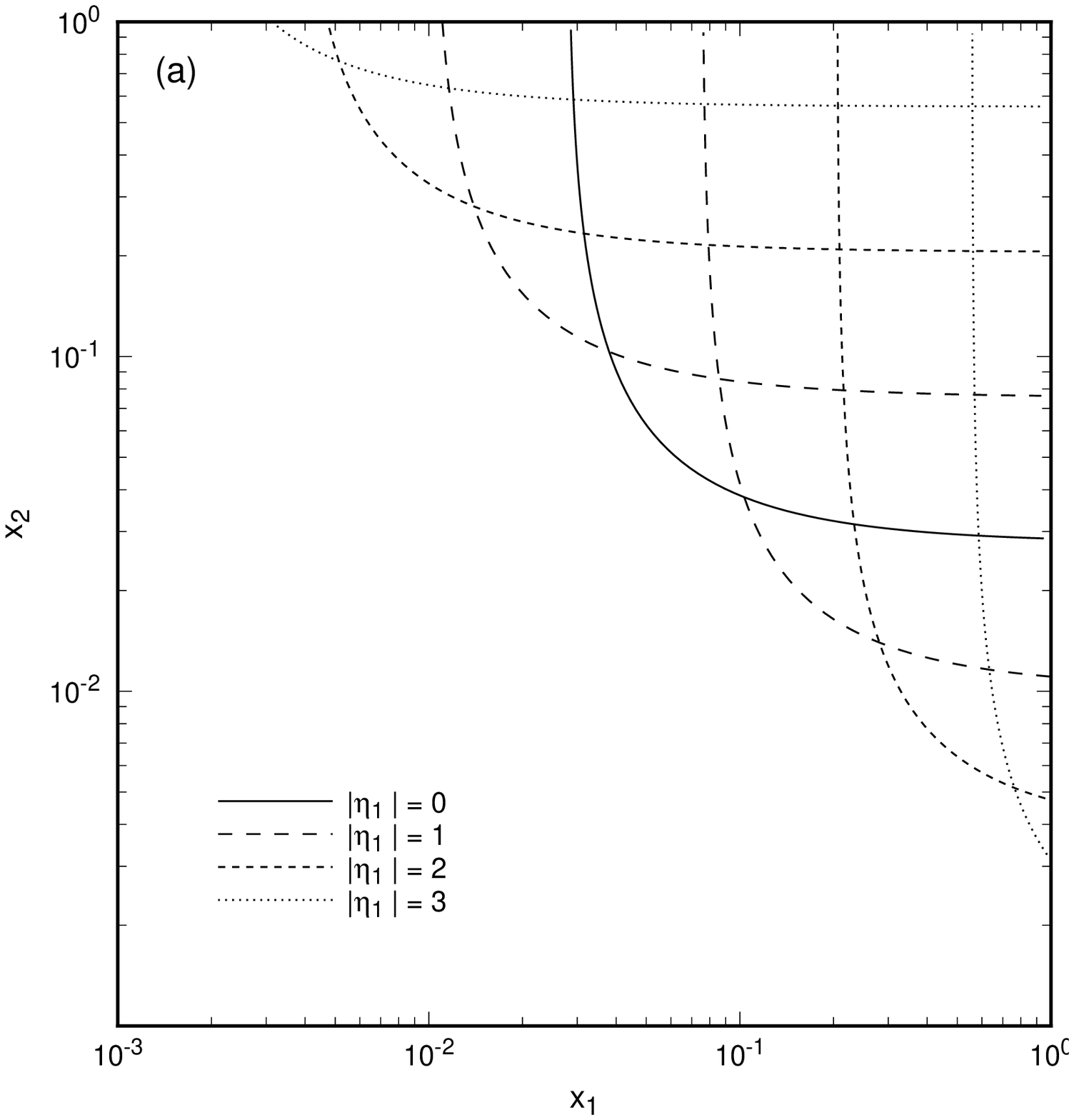}
\includegraphics{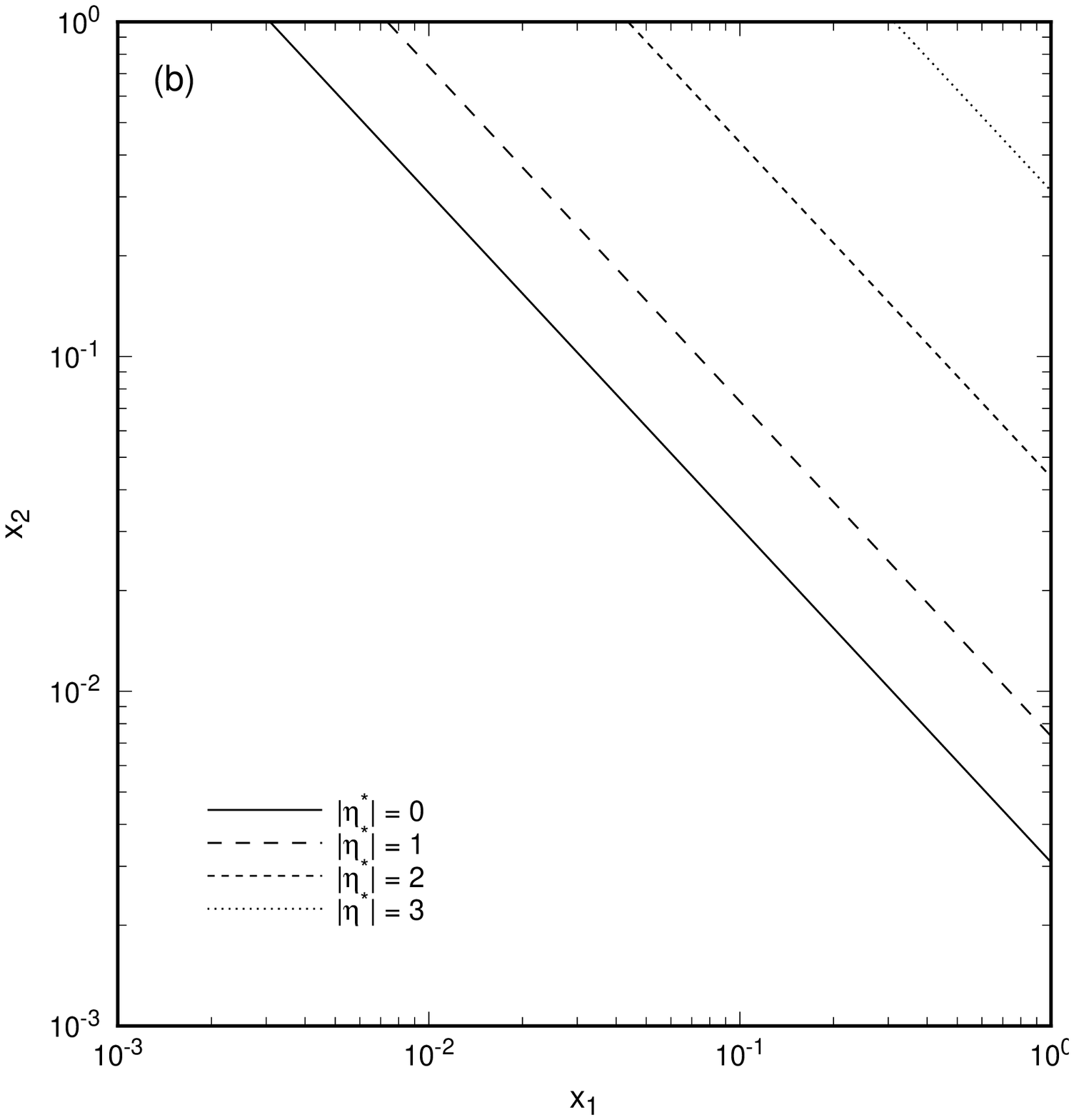}
\caption[]{Contours of (a) constant $\eta_1$ and (b)
constant $\eta^*$ in the $x_1-x_2$ plane for $E_T = 50$~GeV and
$\sqrt{s} = 1800$~GeV.}
\end{figure}

At lowest order in perturbative QCD, 
each point in the $\eta_1-\eta_2$ plane is uniquely
related to the parton momentum fractions $x_1$ and $x_2$ via Eq.~1.
To give an idea of how different $x$ values are spread over the
allowed pseudorapidities, we show contours of fixed $\eta_1$ in the
$x_1-x_2$ plane in Fig.~2.  At fixed $\eta_1$, the smallest $x_1$
value occurs when $\eta_2$ is a minimum, while $\eta_2$ is a maximum
for the smallest $x_2$ value.  An alternative way of looking at the
$x_1-x_2$ plane using the $\eta^*$ and $\etaboost$ variables is
shown in Fig.~2b.  For a given $\eta^*$, varying $\etaboost$ over
its allowed range takes $(x_1,x_2) = (1,x_T^2 \cosh^2(\eta^*))$ to
$(x_1,x_2) = (x_T^2 \cosh^2(\eta^*),1)$ while preserving $x_1 x_2 =
x_T^2 \cosh^2(\eta^*)$.  These contours are particularly useful since
the parton matrix elements depend only on $\eta^*$ and not on
$\etaboost$.

\def\mrsdz{MRSD${}_0$}
\def\mrsdm{MRSD${}_{-}$}
Although the charged parton distributions have been probed directly 
over a wide range of parton momentum fractions $x$ and scales $Q^2$ in
deeply inelastic scattering, the gluon density is rather poorly known.
Direct photon data from WA70 \cite{WA70} determine the shape of the
gluon in the $x \sim 0.3 - 0.4$ region, however, the gluon density at
other $x$ values is only constrained by the momentum sum rule.  To
explore the sensitivity of the triply differential cross section, we
choose parton density functions with contrasting small-$x$ behaviours; the
improved \mrsdm\ and \mrsdz\ distributions of ref.~\cite{MRSD} for which
$xg(x)$ behave as $x^{-0.5}$ and $x^0$ respectively at small $x$ and
$Q^2$.  The low-$x$ behaviour of $F_2^{ep}$ measured at HERA is better
fitted by an $x^{-0.3}$ growth as parameterised by the MRSA
distributions \cite{MRSA}, however the range of predictions from the
\mrsdm\ and \mrsdz\ distributions indicate where the triply differential
cross section is sensitive to the small-$x$ parton distributions.  It
is worth noting that because of the momentum sum rule, a parton
density that is relatively large at $x \sim 10^{-3}$ must be
relatively small at $x \sim {\rm few} \times 10^{-2}$.  This is
demonstrated in Fig.~3a where we show the ratio of the `single
effective parton density' for the \mrsdz\ and \mrsdm\ parton density
functions relative to that of the MRSA parameterisation.  The hierachy
evident at small $x$ is reversed in the moderate $x$ range, where the \mrsdz\
density is 15\% larger than that for the \mrsdm\ density functions.  

At $x \sim 1$, the MRSD
distributions are both larger than MRSA.  This is primarily because
the up and down valence distributions are fitted separately in the
more recent MRS parameterisations rather than the up and the (up plus
down) valence combination.  In any event, the distribution of partons
inside the proton with $x > 0.5$ is very small and poorly constrained
by data.

If we examine the gluon and the quark densities separately (the latter
summed over all quark and anti-quark flavors), shown in
Fig.~3b, then we
find that other than at small~$x$, the quark densities are quite well
determined by present-day data: different sets are quite similar.
In contrast, it is the gluon densities that are poorly determined:
different sets are substantially different even at intermediate~$x$.
As suggested by the `single effective parton density,' and as
we shall see in greater detail, the gluon
densities are sufficiently important to jet production in hadron-hadron
scattering to cause substantial variations in predictions dependent
on the densities at moderate~$x$.

\begin{figure}\vspace{8cm}
\includegraphics{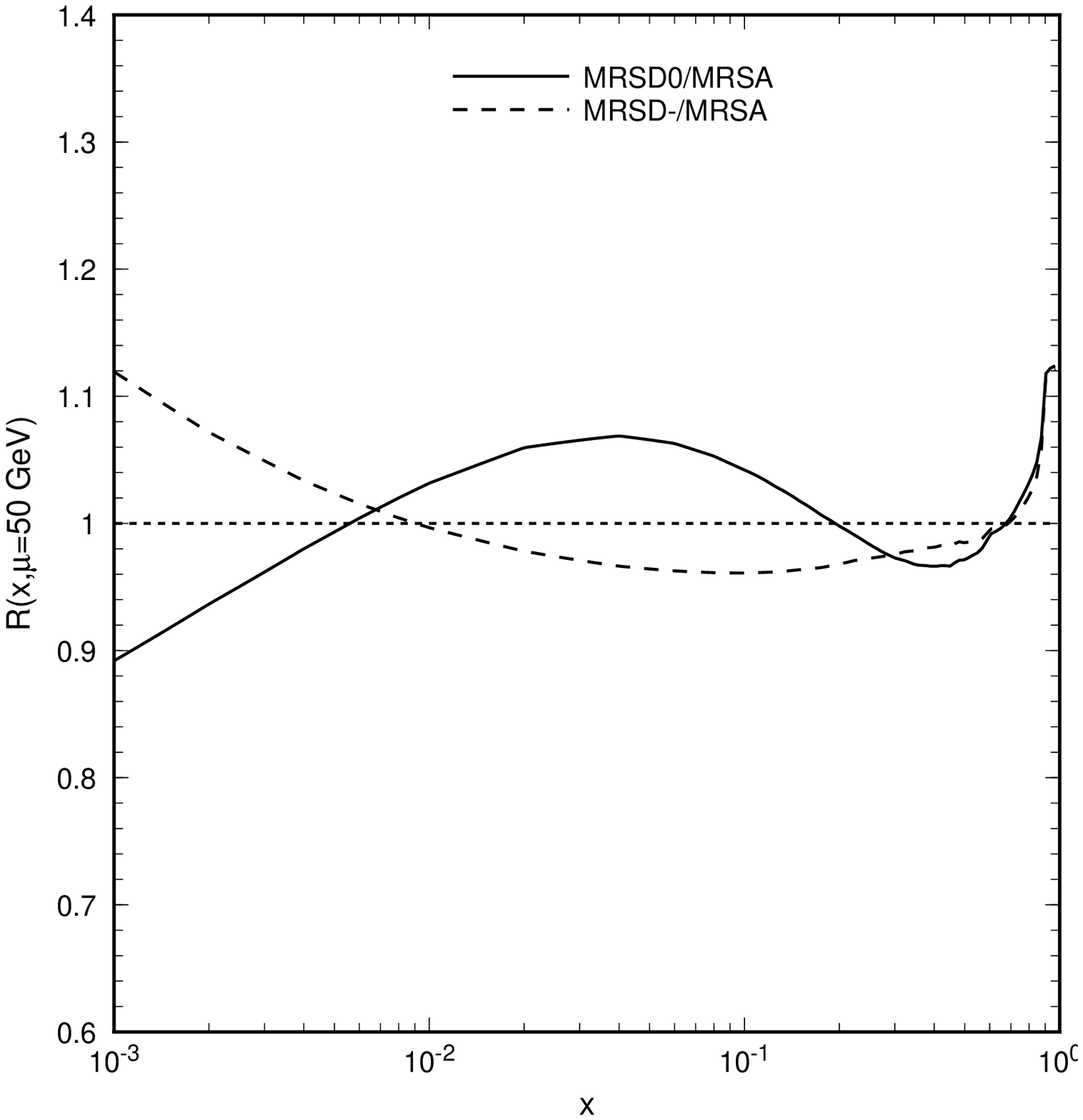}
\includegraphics{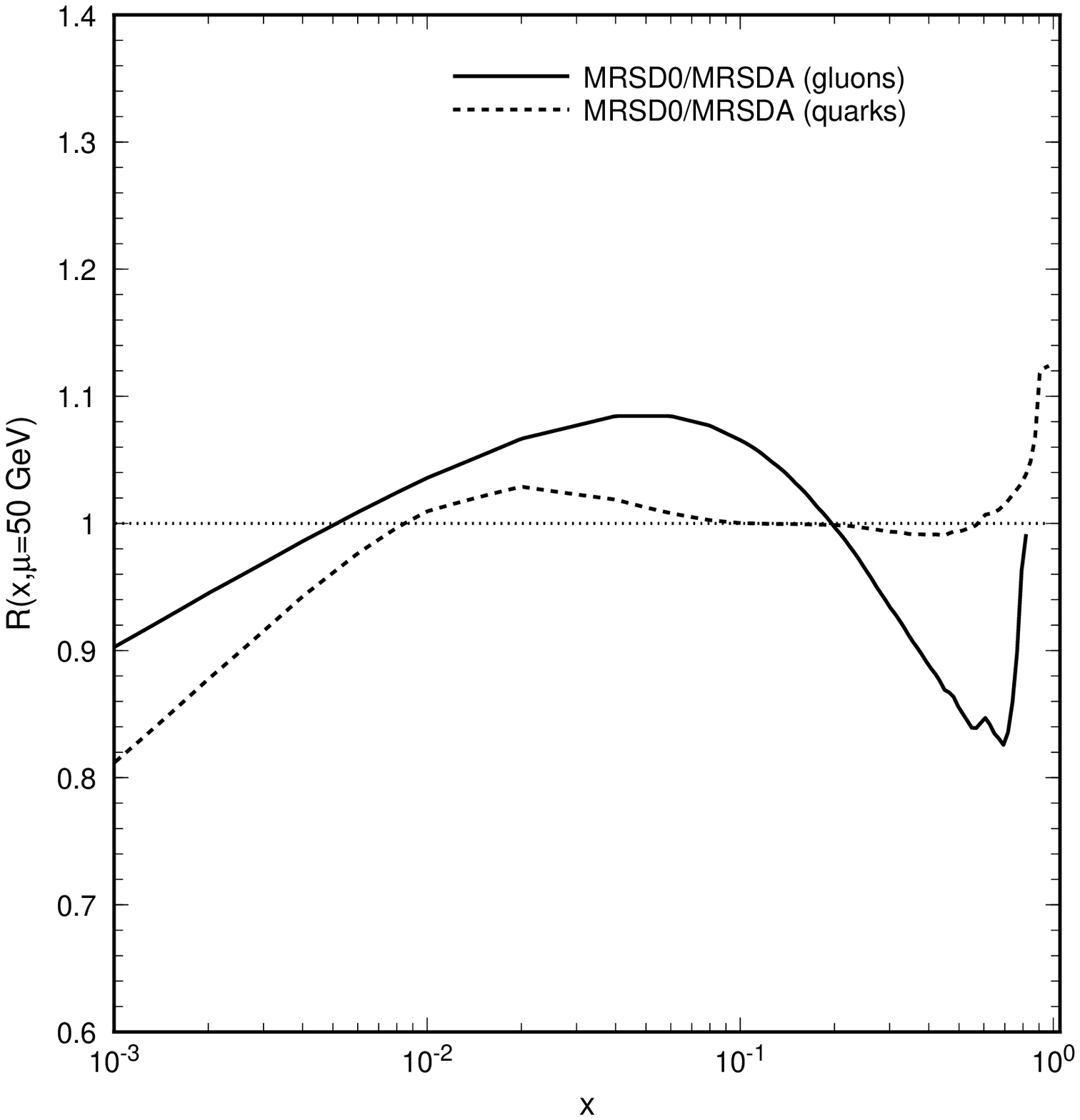}
\caption[]{(a) The ratio of the `single effective parton density' 
of Eq.~4 for the \mrsdz\ and \mrsdm\ distributions compared 
to the MRSA parameterisation at $\mu= 50$~GeV. (b) The ratio of
the gluon and quark parton densities in the \mrsdz\ distribution
compared to the MRSA parameterisation at the same scale.}
\end{figure}

\begin{figure}\vspace{8cm}
\includegraphics{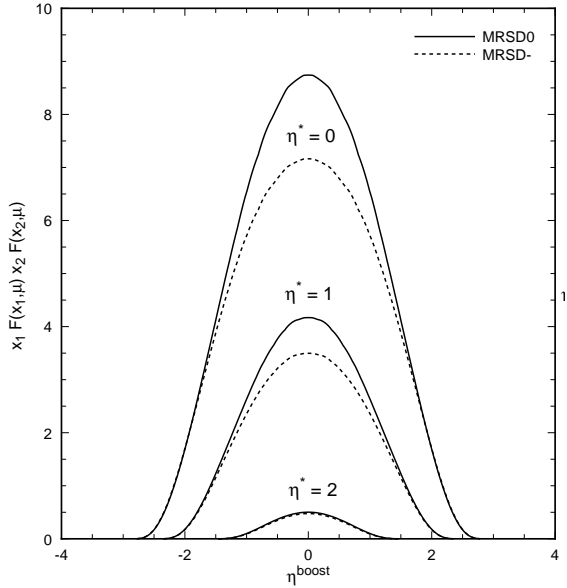}
\caption[]{The parton-parton luminosity for the \mrsdz\ and \mrsdm\ 
parton densities in the `single effective subprocess approximation' as
a function of $\etaboost$ for $|\eta^*| = 0, 1$ and 2 and $\mu =
50$~GeV.}
\end{figure}

As can be seen from Eq.~2, the cross section is proportional to the 
product of structure functions.
To get a feeling for how this product varies,  
Fig.~4 shows the parton-parton luminosity in the single effective
subprocess approximation
as a function of
$\etaboost$ for different $|\eta^*|$ values.  This corresponds to
diagonal strips across the $\eta_1-\eta_2$ plane.  As expected, the
largest luminosity occurs when $x_1$ and $x_2$ are equally small,
$\eta^*\sim \etaboost = 0$.  Once again, the \mrsdz\ luminosity is
approximately 20\% larger than that for \mrsdm\ .  As either
$|\etaboost|$ or $|\eta^*|$ increases, the luminosity decreases
rapidly.  However the falloff is more rapid with increasing $|\eta^*|$
than with increasing $|\etaboost|$.

\begin{figure}\vspace{8cm}
\includegraphics{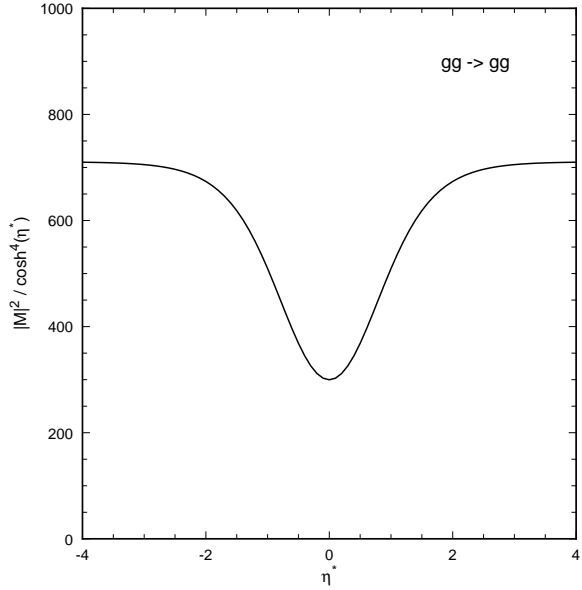}
\caption[]{The lowest order
squared matrix elements $|{\cal M}_{gg}|^2/\cosh^4(\eta^*)$ for $gg\to
gg$ scattering as a function of $\eta^*$.}
\end{figure}

As mentioned earlier, the parton-parton subprocess scattering matrix
elements are independent of $\etaboost$.  One consequence 
is that for fixed $\eta^*$, the only variation of the cross section
comes from the variation of the parton densities as $\etaboost$ runs over the
allowed kinematic range.  The lowest order squared matrix elements for
$gg \to gg$ scattering are given by,
\begin{equation}
\frac{|{\cal M}_{gg}|^2}{\cosh^4(\eta^*)}
  = \frac{9\pi^2}{8}  \frac{(4\cosh^2(\eta^*)-1)^3}{\cosh^6(\eta^*)}.
\end{equation}
These matrix elements are plotted in Fig.~5 as a function of $\eta^*$.
At small $|\eta^*|$ the matrix elements grow rapidly until
$|\eta^*|\sim 2$ where the matrix elements saturate.  This behaviour
complements the parton-parton luminosity which is largest at $\eta^* =
0$.

By multiplying the parton-parton luminosity with the squared matrix
elements (along with the overall factor $\alpha_s^2(\mu)/8\pi E_T^3$ )
we obtain the physical cross section.  Because different parton
distributions dominate for different momentum fractions (and hence
$\eta_1$ and $\eta_2$ values), we expect the shape of the 
triply-differential cross section to be sensitive to the parton densities.  
This is illustrated in Fig.~6, where we show the leading order prediction
for the ratio,
\begin{equation}
R = \frac{
\int^{E_{T{\rm max}}}_{E_{T{\rm min}}} 
dE_T \left (\frac{d^3\sigma}{dE_Td\eta_1d\eta_2}({\rm MRSD}_0)   
-\frac{d^3\sigma}{dE_Td\eta_1d\eta_2}({\rm MRSD}_{-})  \right) }
{\int^{E_{T{\rm max}}}_{E_{T{\rm min}}} 
dE_T \frac{d^3\sigma}{dE_Td\eta_1d\eta_2}({\rm MRSD}_{-})   },
\end{equation}
in the transverse energy range 45~GeV $< E_T < $ 55~GeV evaluated at
$\mu = E_T$\footnote{The raw cross sections are histogrammed in $0.5
\times 0.5$ bins in $\eta_1$ and $\eta_2$.}.  In addition to the sharp
cutoff marking the boundary of the allowed phase space, the excess of
\mrsdz\ over \mrsdm\ at small $\eta_1$ and $\eta_2$ and the depletion at
large $|\eta_1| \sim |\eta_2|$ are seen clearly.  Over the whole
$\eta_1-\eta_2$ plane, the relative cross sections vary by $+23\%$ to
$-5\%$, with the most sizeable effects at $\eta_1 \sim \eta_2 \sim 0$
where the cross section is largest.  At lowest order, $E_{T1} = E_{T2}
= E_T$, so that the distribution is symmetric under $\eta_1
\leftrightarrow \eta_2$.

\begin{figure}\vspace{8cm}
\includegraphics{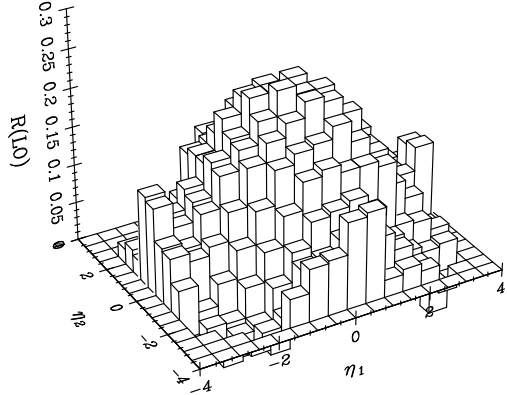}
\caption[]{The leading order prediction for the
ratio of the triply differential cross section $R(LO)$ defined in
Eq.~8 for 45~GeV $ < E_T < 55$~GeV and $\mu = E_{T}$.}
\end{figure}

\section{The ${\cal O}(\alpha_s^3)$ triply differential two jet
cross section}

At next-to-leading order, some contributions admit
three partons into the final
state; for these, the parton fractions are given by,
\begin{equation}
x_{1,2} = \frac{E_{T1}}{\sqrt{s}}
\left (\exp(\pm\eta_1) 
+ \frac{E_{T2}}{E_{T1}}\exp(\pm\eta_2)
+ \frac{E_{T3}}{E_{T1}}\exp(\pm\eta_3)
\right),
\end{equation}
where $E_{Ti}$ and $\eta_{i}$ $(i=1,\ldots,3)$ describe the transverse
energies and pseudorapidities of the three partons ordered in
decreasing $E_T$.  Since the transverse energies of the partons are no
longer forced to be equal, $|\eta_2|$ may increase to compensate for
having a smaller transverse energy, $E_{T2}/E_{T1} < 1$.  The maximum
possible values of $|\eta_2|$ occur when $E_{T2} = E_{T3}$,
\begin{equation}
-\log\left(\frac{a\bar a +\sqrt{a^2\bar a^2-a\bar a}}{a}\right ) < \eta_2 < 
\log\left(\frac{a\bar a +\sqrt{a^2\bar a^2-a\bar a}}{\bar a}\right ),
\end{equation}
where $a = (2-x_{T1}\exp(\eta_1))/x_{T1}$ and $\bar a =
(2-x_{T1}\exp(-\eta_1))/x_{T1}$.  The enlargement of phase space is
shown in Fig.~1 (region III).  We see that the maximum allowed value
of $|\eta_1|$ is unchanged at next-to-leading order so that the
physical cross section will exhibit a rather sharp cutoff as
$|\eta_1|$ increases.  On the other hand there will be a more gradual
fall off in the cross section as $|\eta_2|$ increases.  Indeed,
adding more partons into the final state further increases the allowed
$\eta_2$ range corresponding to the production of more and more soft
partons.

To compute the next-to-leading order cross section, we use an ${\cal
O}(\alpha_s^3)$ Monte Carlo program for one, two and three jet
production based on the one-loop $2 \to 2$ and the tree level $2 \to
3$ parton scattering amplitudes \cite{ES,OtherFour} described in
ref.~\cite{GGK2jet}.  This program uses the techniques of
refs.~\cite{GG,GGK} to cancel the infrared and ultraviolet
singularities thereby rendering the $2\to 2$ and $2\to 3$ parton
processes finite and amenable to numerical computation.  The parton
four momenta are then passed through a jet algorithm to determine the
one, two and three jet cross sections according to the experimental cuts.
Different cuts and/or jet algorithms can easily be applied to the
parton four-momenta and, in principle, any infrared-safe distribution
can be computed at ${\cal O}(\alpha_s^3)$.

In order to compare the theory with experiment, we use the parton
level equivalent of the standard `Snowmass' cone algorithm~\cite{Snowmass} with
$\Delta R = 0.7$ and require at least two jets in the event.
Furthermore, we note that the assignment of which jet is hardest is
not infrared safe, so that we must symmetrize over the hardest and
next-hardest jets (in transverse energy).  The distributions we shall
examine require that the hardest jet lie in a certain `trigger' range;
each event will be counted twice if the next-hardest jet also lies
in this $E_T$ range. The three-dimensional
cross section evaluated at $\mu = E_{T1}$, where $E_{T1}$ is the
transverse energy of the hardest jet in the event, is shown in Fig.~7
for 45~GeV $< E_T < 55$~GeV.  The extension of the phase space to
smaller $\eta_2$ is seen clearly, along with the rather sharp cutoff
at $|\eta_1|\sim 3.5$.

\begin{figure}
\vspace{8cm}
\includegraphics{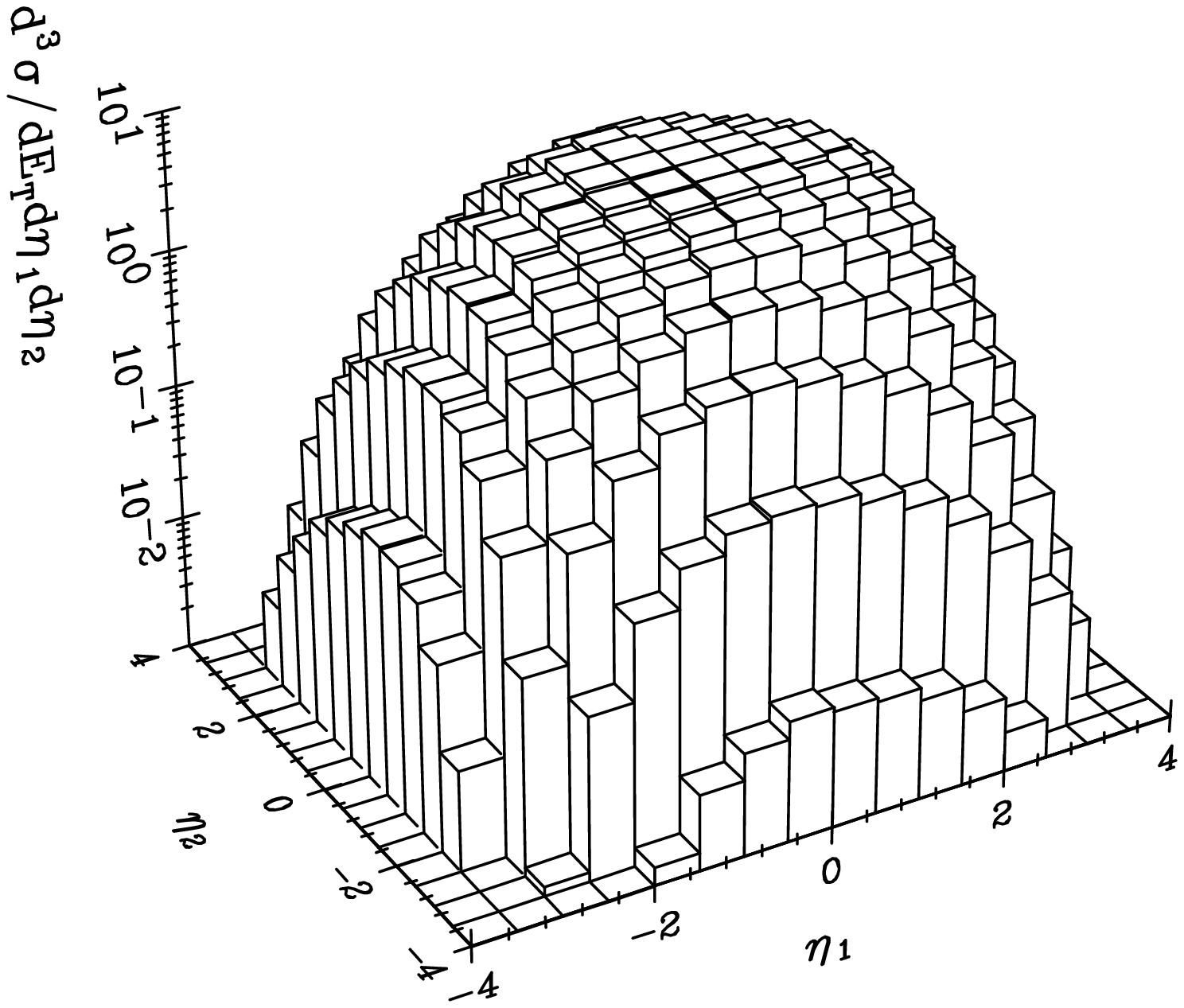}
\includegraphics{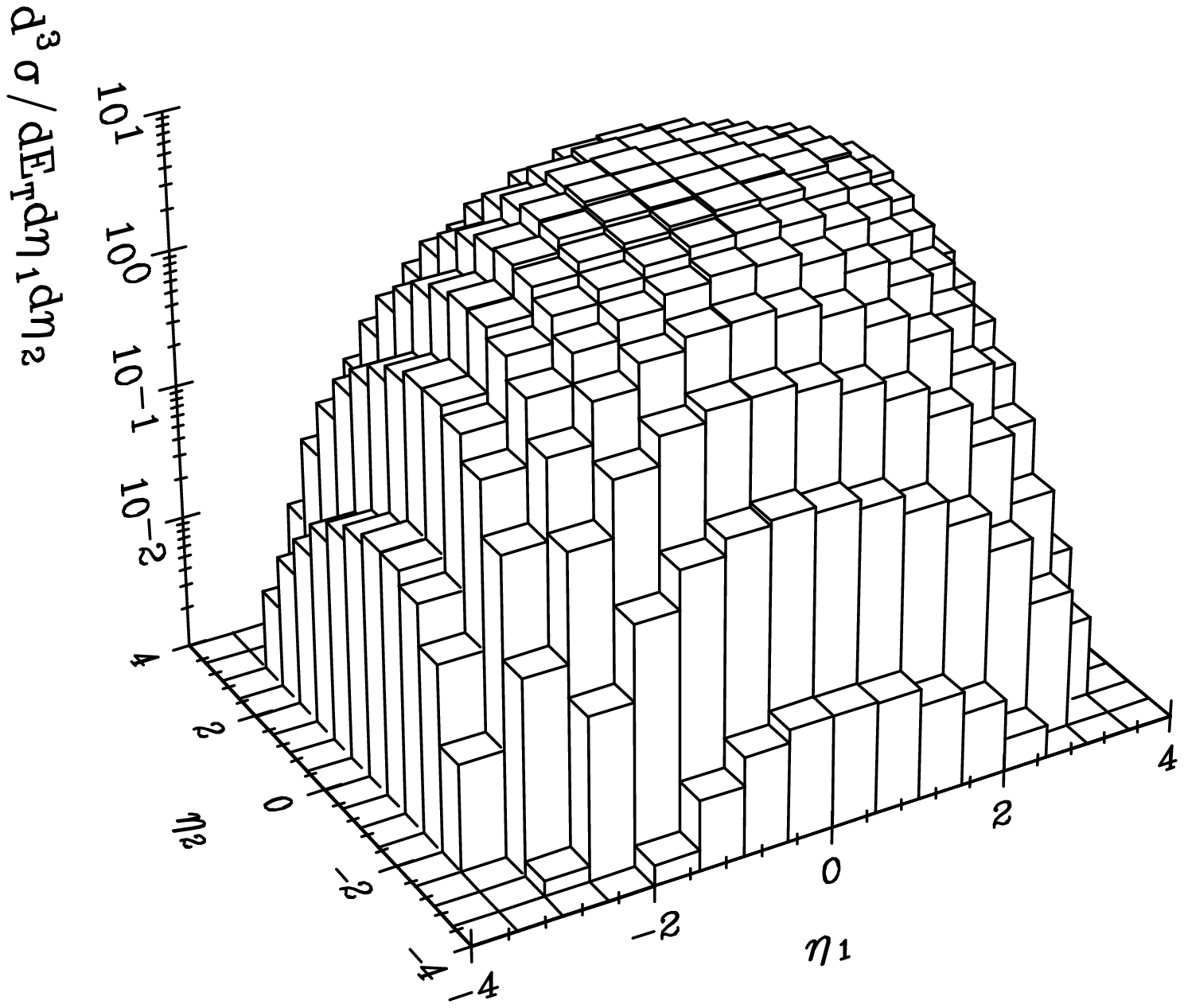}
\vspace{-1cm}\hspace{2cm}(a)\hspace{7cm}(b)\vspace{1cm}
\caption[]{The next-to-leading order triple differential distribution
for 45~GeV $ < E_T < 55$~GeV and $\mu = E_{T1}$ for (a) \mrsdz\ and (b)
\mrsdm\ parton densities.}
\end{figure}

Although the cross sections for the two parton densities
appear similar, the difference between the
predictions observed at lowest order is
preserved.  This is illustrated in Fig.~8, where we show 
the next-to-leading order prediction for the fractional difference ratio
defined in Eq.~8.  As at lowest order, the difference varies between
$+23\%$ at $\eta_1 \sim \eta_2 \sim 0$ and $-10\%$ at $\eta_1 \sim
\eta_2 \sim -2.5$.  We note that the ratio is most negative when $x_1
\sim x_T^2$ and $x_2 \sim 1$.  
This is the region where the singular behaviour of the \mrsdm\ parton densities
dominate over the less singular \mrsdz\ distributions~\cite{GGKssos}.

\begin{figure}\vspace{8cm}
\includegraphics{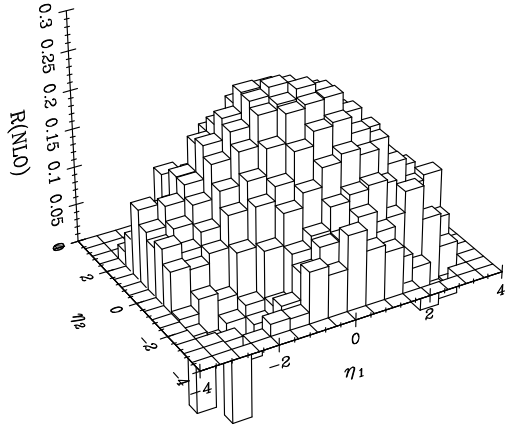}
\caption[]{The next-to-leading order prediction for the
ratio of the triply differential cross section $R(NLO)$ defined in
Eq.~8 for 45~GeV $ < E_T < 55$~GeV and $\mu = E_{T1}$.}
\end{figure}

One indicator of the reliability of perturbation theory is the ratio
of next-to-leading order to leading order cross sections\footnote{The
normalisation is extremely sensitive to the renormalization scale 
choice in the leading order cross section.}.  This is shown in Fig.~9 for 
two slices
of the triply differential distribution.  First we consider the slice
$\eta_1 = 0$ and let $\eta_2$ vary.  At large $|\eta_2|$, the phase
space extends beyond the strict leading-order kinematic limit of
$\eta_2 < \cosh^{-1}(1/x_T) = 3.68$.
As a consequence, the corrections are large.  However, this does {\it
not\/} signal the emergence of large logarithms which might spoil the
applicability of perturbation theory.  Rather, the large corrections
are due to the more restricted phase space available at leading order,
that is the absence of region~III of Fig.~1.  Towards the edges of
available phase space, the leading-order cross section is thus forced
artificially to zero, and the ratio goes to infinity.  At yet-higher
order, however, one expects further corrections to be reasonable (until
one approaches the outer edges of region~III or its higher-order analogs).
Second, we keep $\eta_2 = 0$
fixed and allow $\eta_1$ to vary.  We see that the corrections for
central $|\eta_1|$ are small, however as the magnitude of $\eta_1$ approaches
the edge of phase space, the next-to-leading order corrections
significantly reduce the cross section.  In this limit, the second jet
is forced to have $E_{T2} \sim E_{T1}$ and the available phase space
for soft gluon emission is curtailed.  As a consequence, radiative
corrections lower the cross section close to the edge of phase space.
These corrections are a result of the appearance of large logarithms,
and one does expect perturbation theory to behave badly near this
edge of phase space.
In summary, we see that the triply differential cross section is reliably
predicted over the whole range of the $\eta_1 - \eta_2$ plane with the
exception of the very large $|\eta_1|$ slices.

\begin{figure}\vspace{8cm}
\includegraphics{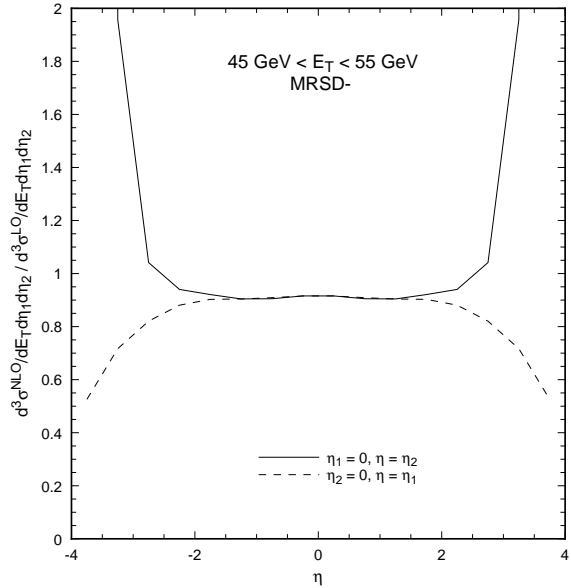}
\caption[]{The ratio of next-to-leading order (NLO) to leading order (LO) 
predictions as a function of $\eta_2$ for $\eta_1 = 0$ (solid) and of
$\eta_1$ for $\eta_2=0$ (dashed) for 45~GeV $< E_T < $~55~GeV, $\mu =
E_T$ and the \mrsdm\ structure functions.}
\end{figure}

We have seen how the shape of the two-dimensional distribution is
sensitive to the parton density functions.  However, there is also a
dependence on the renormalisation and factorisation scales $\mu_R$ and
$\mu_F$\footnote{Throughout we choose $\mu_R = \mu_F = \mu$, however,
other choices are possible.} that could in principle obscure the
differences due to the structure functions.  To get a feeling of how
severely the scale uncertainty affects the shape, Fig.~10 shows the
next-to-leading order predictions for the ratio,
\begin{equation}
R = \frac{
\int^{E_{T{\rm max}}}_{E_{T{\rm min}}} dE_T 
\left (\frac{d^3\sigma}{dE_Td\eta_1d\eta_2}(\mu = E_{T1})   
-c   
\frac{d^3\sigma}{dE_Td\eta_1d\eta_2}
(\mu = \lambda E_{T1}) \right )  }
{\int^{E_{T{\rm max}}}_{E_{T{\rm min}}} dE_T 
\frac{d^3\sigma}{dE_Td\eta_1d\eta_2}
(\mu = E_{T1})   },
\end{equation}
for the \mrsdm\ parton densities and $\lambda= 0.5$ and 2.  Because the
absolute magnitude of the
cross section (which depends on $\alpha_s(\mu)$) is poorly predicted,
the prediction for $\mu = \lambda E_{T1}$ has been normalised to the
cross section for $\mu = E_{T1}$ at $\eta_1 \sim \eta_2 \sim 0$.  For
$\lambda= 0.5$, $c=0.93$, while for $\lambda = 2$, $c=1.08$.
We have restricted the pseudorapidity range in the plot 
to $|\eta_1|,~|\eta_2| < 2.5$
since for higher pseudorapidities 
the next-to-leading order effects are large as discussed
above. As a result, for such pseudorapidities, there is a sizeable scale  
variation.
However,  for central pseudorapidities, the shape is changed by less than 5\%
which is significantly less than the difference between the two
representative parton distributions.

\begin{figure}
\vspace{8cm}
\includegraphics{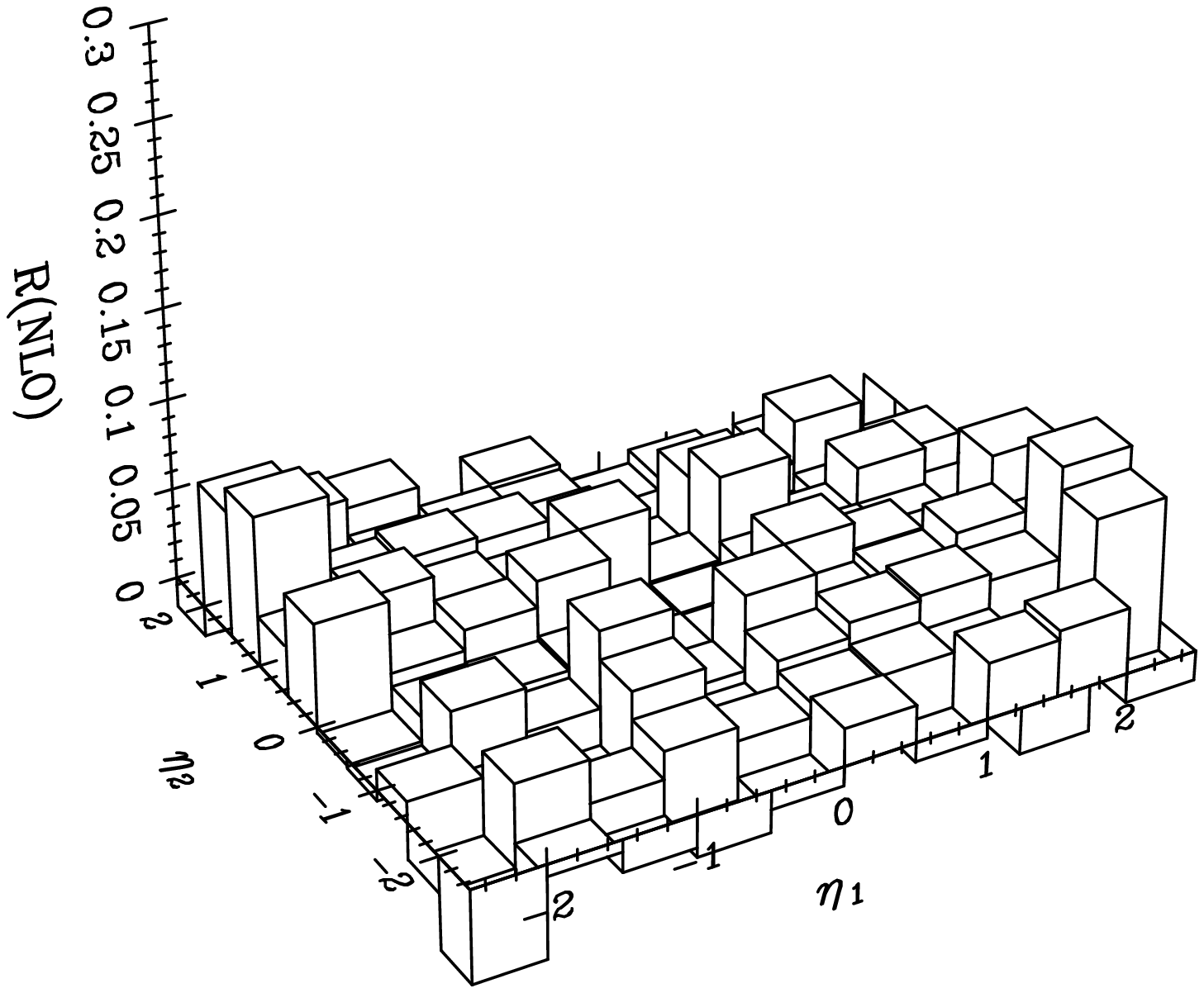}
\includegraphics{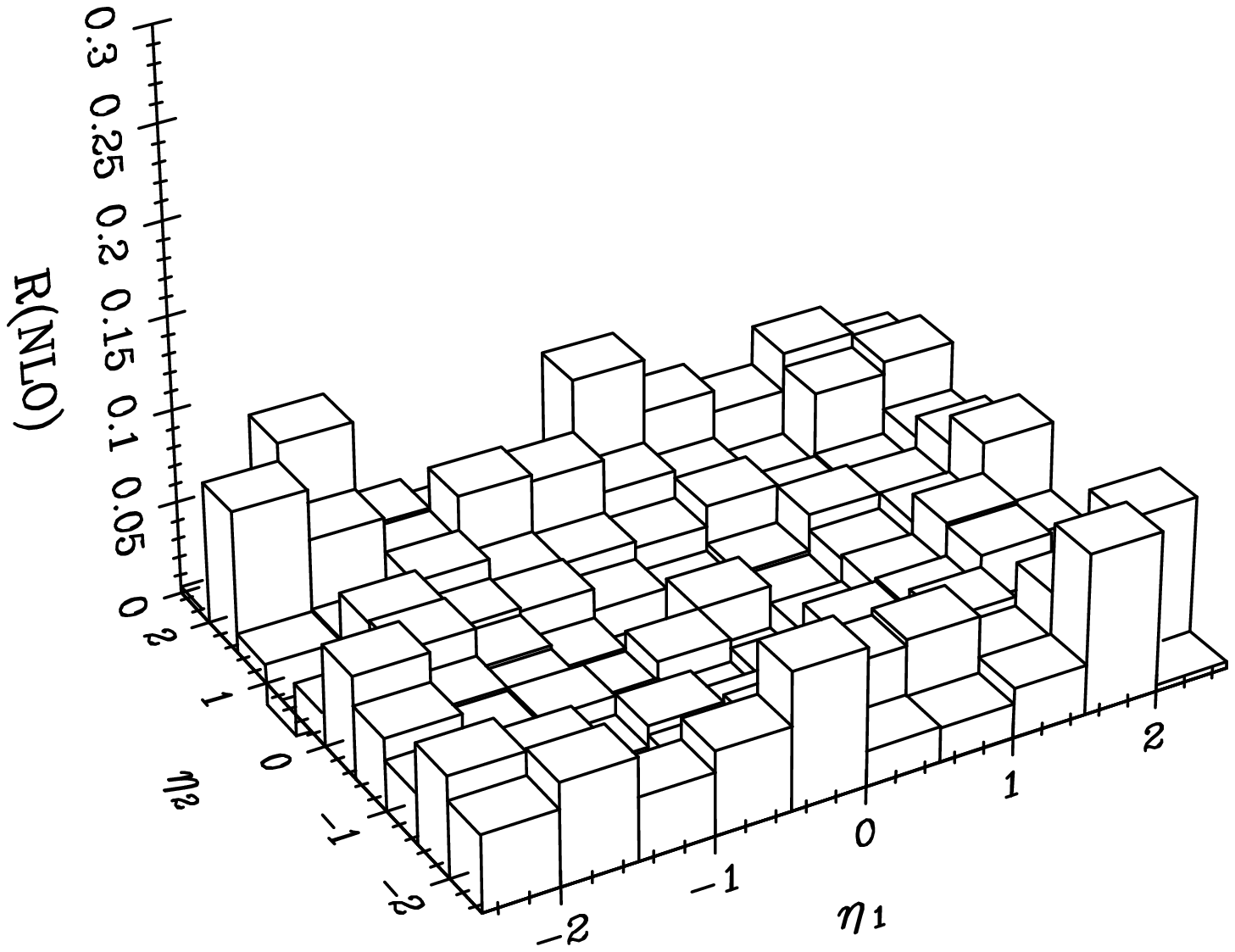}
\vspace{-1cm}\hspace{2cm}(a)\hspace{7cm}(b)\vspace{1cm}
\caption[]{The next-to-leading order prediction for the
ratio of the triply differential cross section $R(NLO)$ 
as defined in Eq.~11 with (a) $\lambda=0.5$ and (b) $\lambda=2$ 
for 45~GeV $ < E_T < 55$~GeV and $\mu = E_{T1}$.}
\end{figure}

Once the experimental data are available, it should therefore be
possible to extract information on the density of partons in the
proton.  In addition to the uncertainty in the normalisation of the
theoretical predictions, there is a significant uncertainty in the
experimental normalisation as well,
due to uncertainties in the luminosity measurement, jet
energy calibration, jet trigger efficiency,
and other aspects\footnote{Part of the experimental
uncertainty can be eliminated by normalising with respect to the $W$
cross section.}.  It therefore makes sense to allow the overall
normalisation of the theoretical prediction, $\sigma^{TH}$, to float,
so that by varying $c$ the $\chi^2$ for
\begin{equation}
\int^{E_{T{\rm max}}}_{E_{T{\rm min}}}  dE_T
\left (\frac{d^{3}\sigma^{EXP}}{dE_Td\eta_1d\eta_2} - c  
\frac{d^{3}\sigma^{TH}}{dE_Td\eta_1d\eta_2}\right),
\end{equation}
summed over the different $\eta_1$, $\eta_2$ cells is minimised for a
given parameterisation of the parton densities.   Finally,
the input parameterisations can be adjusted so that the $\chi^2$ is
further reduced. This can be done
simultaneously for different slices in transverse energy.
An even more interesting possibility would be to map out the evolution 
of the parton densities directly by following 
trajectories of constant $(x_1,x_2)$ in the
$\eta_1-\eta_2$ plane as a function of $E_T$.

\section{The D0 signed distribution}

Recently, the D0 collaboration has presented preliminary 
data \cite{Weerts,D0sign} for
a particular slicing of the triply differential distribution --- the 
so-called signed pseudorapidity distribution.  This amounts to taking two
strips of the $\eta_1-\eta_2$ plane for a fixed transverse energy
interval and combining them in reverse directions.  The pseudorapidity of
the leading jet is constrained to lie in the range $|\eta_1|_{\rm min}
< |\eta_1| < |\eta_1|_{\rm max}$ and the distribution is plotted as a
function of $|\eta_2| {\rm sign}(\eta_1 \eta_2)$,
\begin{equation}
\hskip -10pt\frac{d\sigma}{d|\eta_2|{\rm sign}(\eta_1 \eta_2)}
\equiv {1\over\Delta E_T} \int^{E_{T{\rm max}}}_{E_{T{\rm min}}} dE_T\,
{1\over 2\Delta \eta_1}
\left ( \int^{|\eta_1|_{\rm max}}_{|\eta_1|_{\rm min}} d\eta_1 
 \frac{d^3\sigma}{dE_Td\eta_1d\eta_2} -
\int^{-|\eta_1|_{\rm min}}_{-|\eta_1|_{\rm max}} d\eta_1 
\frac{d^3\sigma}{dE_Td\eta_1d\eta_2} \right ),
\end{equation}
where ${\rm sign}(\eta_1 \eta_2) = -1$ if $\eta_1$ and $\eta_2$ have
opposite sign and +1 if they have the same sign.  Positive values of
$|\eta_2| {\rm sign}(\eta_1 \eta_2)$ correspond to same-side dijet
events, while negative values are associated with opposite-side
events.  In principle, both strips contain equal information, but
combining them serves to reduce the statistical error.  Once again,
we must sum over the hardest and next-hardest jet in order to
ensure that this distribution is infrared-safe.

In the currently available data, D0 has examined two slices in
transverse energy, 45~GeV $< E_T < $ 55~GeV and 55~GeV $< E_T < $
65~GeV, and two strips in $\eta_1$, $0.0 < |\eta_1| < 0.5$ and $2.0 <
|\eta_1| < 2.5$.  As more data become available from the current
Tevatron run, this analysis can be extended to cover a larger range of
$E_T$ and $\eta$.

\begin{figure}\vspace{8cm}
\includegraphics{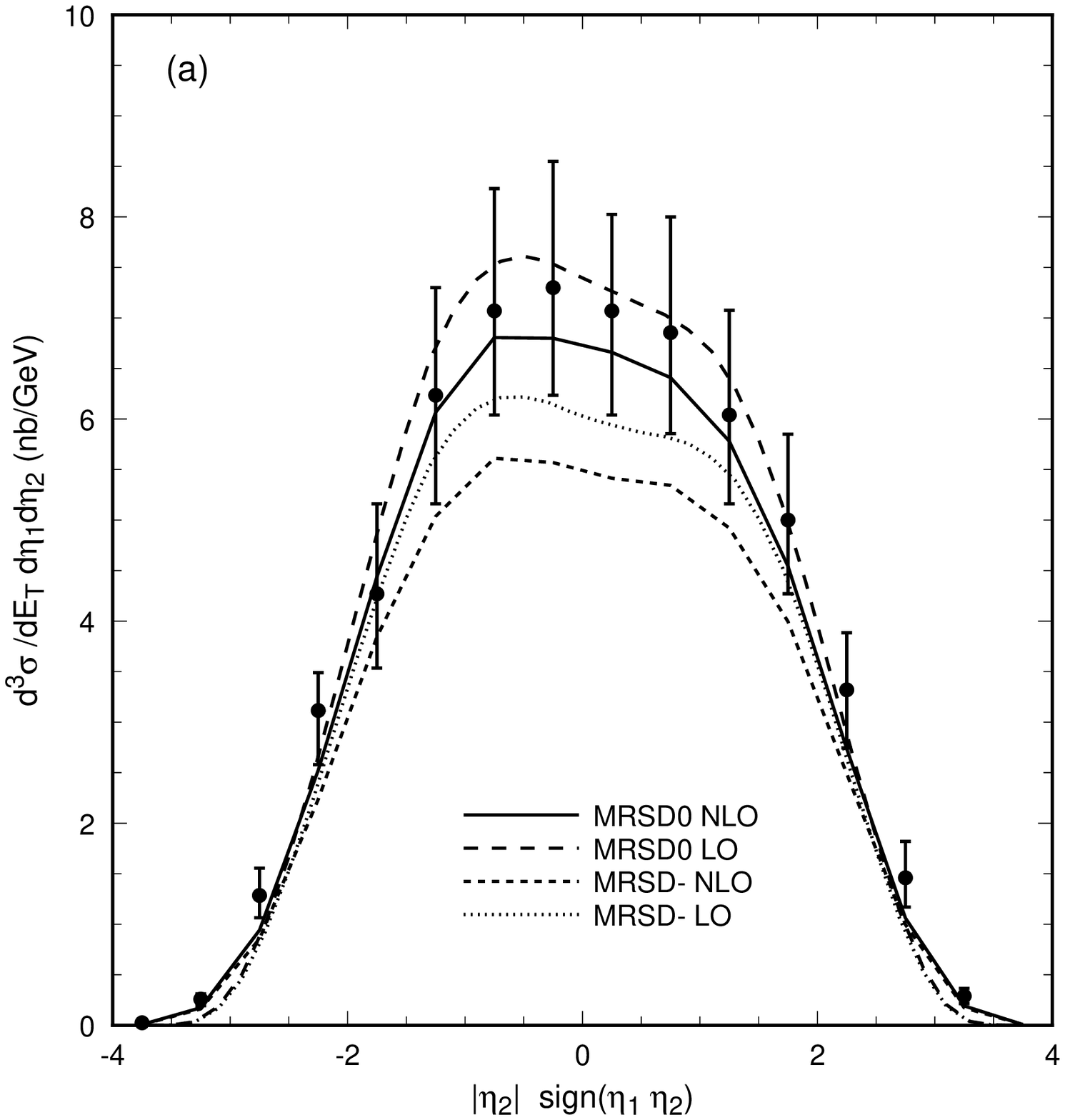}
\includegraphics{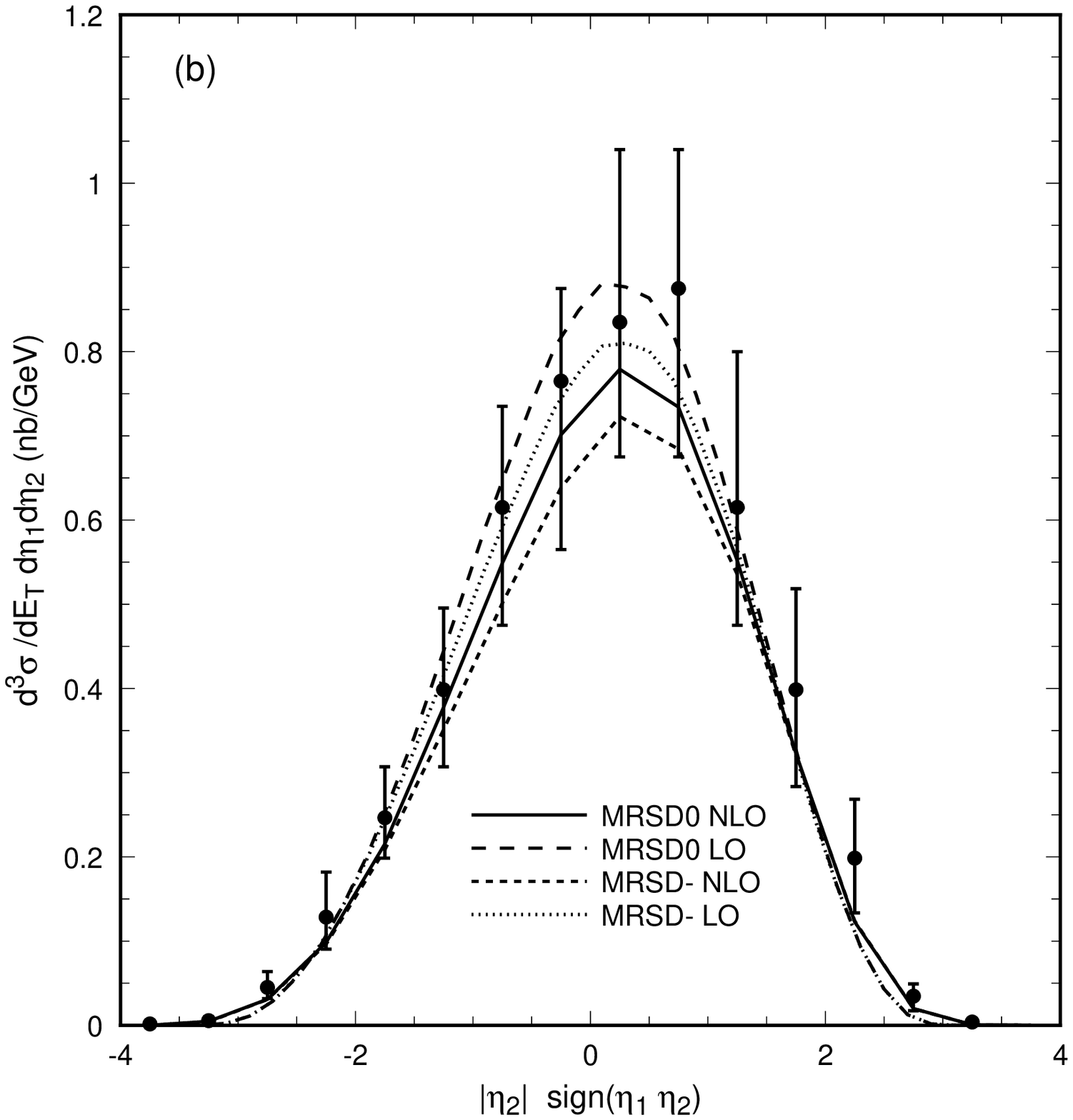}
\caption[]{The signed pseudorapidity distribution for 
(a) 45 GeV $< E_T <$ 55 GeV, 0.0 $< |\eta_1| <$ 0.5 and (b) 55 GeV $<
E_T <$ 65 GeV, 2.0 $< |\eta_1| <$ 2.5 at both LO and NLO. The
preliminary experimental results from \cite{Weerts} are also shown.
The factorisation scale is chosen to be the transverse energy of the
hardest jet, $\mu = E_{T1}$.}
\end{figure}

We first fix $\eta_1$ to lie in the central pseudorapidity slice, $0.0 <
|\eta_1| < 0.5$  and examine the pseudorapidity
of the second jet for the transverse energy interval
45~GeV $< E_T < $ 55~GeV.  
This strip includes the $\eta_1\sim \eta_2 \sim 0$
region that is sensitive to parton densities at $x \sim 0.05$.
We also consider the slice at larger $\eta_1$, 
$2.0 < |\eta_1| < 2.5$  but for a slightly higher transverse energy interval,
55~GeV $< E_T < $ 65~GeV.  The predictions for these distributions
for both \mrsdz\ and \mrsdm\ parton densities are shown in Fig.~11
with the preliminary data from the D0 collaboration 
\cite{Weerts}\footnote{We have divided the data by a
factor of two to account for the size of the pseudorapidity interval.}.  
We see a clear asymmetry favouring smaller values of
$|\eta_2| {\rm sign}(\eta_1 \eta_2)$.  In other words, for $\eta_1
\sim 0$, the opposite-side cross section
(negative $|\eta_2| {\rm sign}(\eta_1 \eta_2)$)
 peaks away from $\eta_2\sim 0$, while the same-side cross section 
(positive $|\eta_2| {\rm sign}(\eta_1 \eta_2)$)
monotonically decreases.  A similar
effect has been observed in the same-side/opposite-side cross section
measured by CDF \cite{CDFssos}. 


This is
is due to an interplay between the parton-parton 
luminosity and the matrix elements. Fig.~12 shows  the parton-parton
luminosity in the `single effective subprocess approximation'
and the $gg\to gg$ matrix elements of Eq.~7 as a function of
$|\eta_2| {\rm sign}(\eta_1 \eta_2)$.  We see
that the maximum of the parton-parton luminosity occurs at 
$|\eta_2| {\rm sign}(\eta_1 \eta_2) \sim
0.25$ and $0.75$ for $\eta_1= 0.5$ and 2.5 respectively, while the
minimum of the matrix elements always lies at 
$|\eta_2| {\rm sign}(\eta_1 \eta_2) = |\eta_1|$.
The net effect of the
shift in the peak of the parton-parton luminosity to positive 
$|\eta_2| {\rm sign}(\eta_1 \eta_2)$ combined with the shift of the
minimum of the matrix elements to larger values of 
$|\eta_2| {\rm sign}(\eta_1 \eta_2)$ is an enhancement 
of the cross section at negative
$|\eta_2| {\rm sign}(\eta_1 \eta_2)$
and a depletion at positive $|\eta_2| {\rm sign}(\eta_1 \eta_2)$,
clearly visible as an asymmetry in Fig.~12.
\begin{figure}\vspace{9cm}
\includegraphics{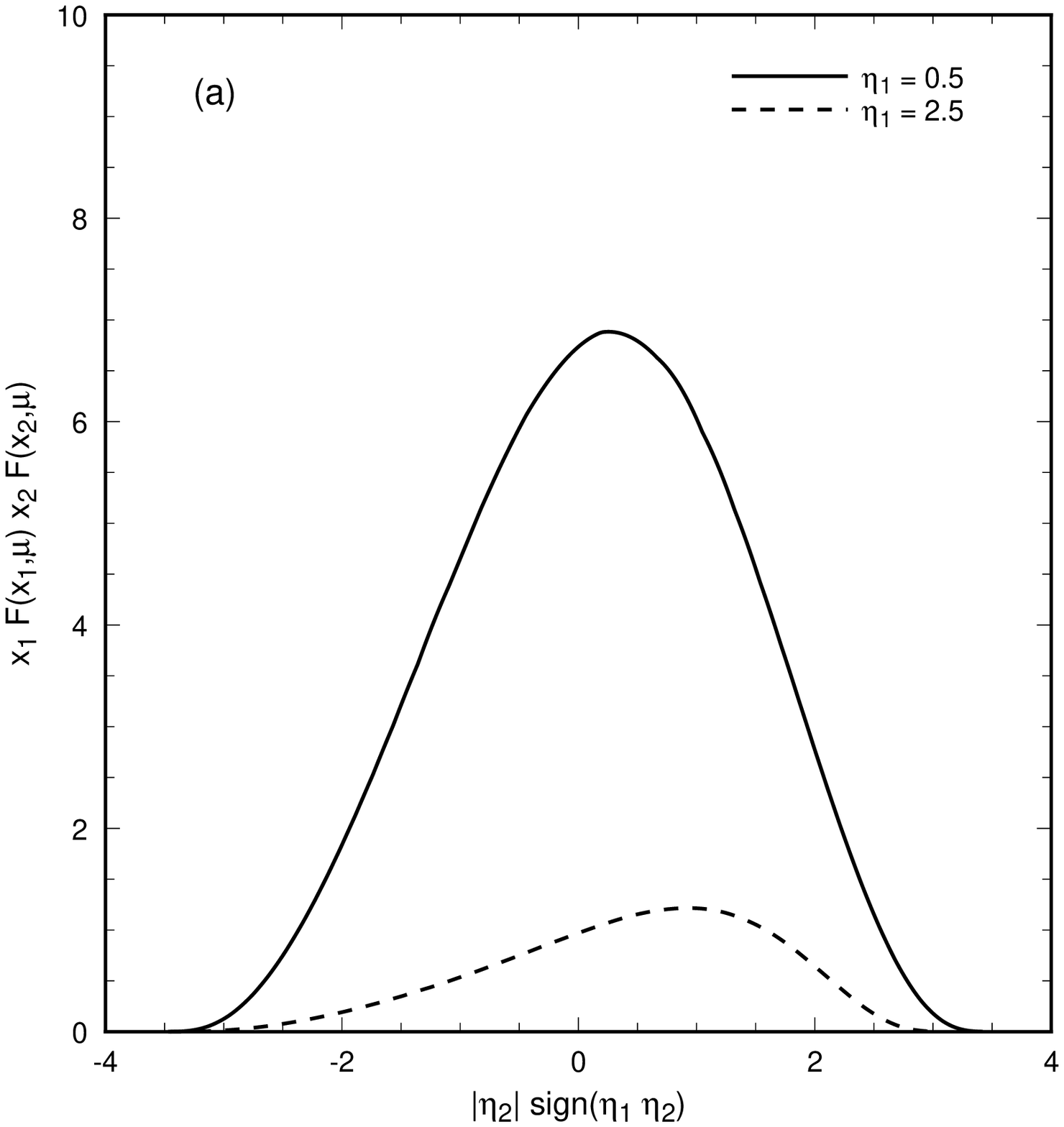}
\includegraphics{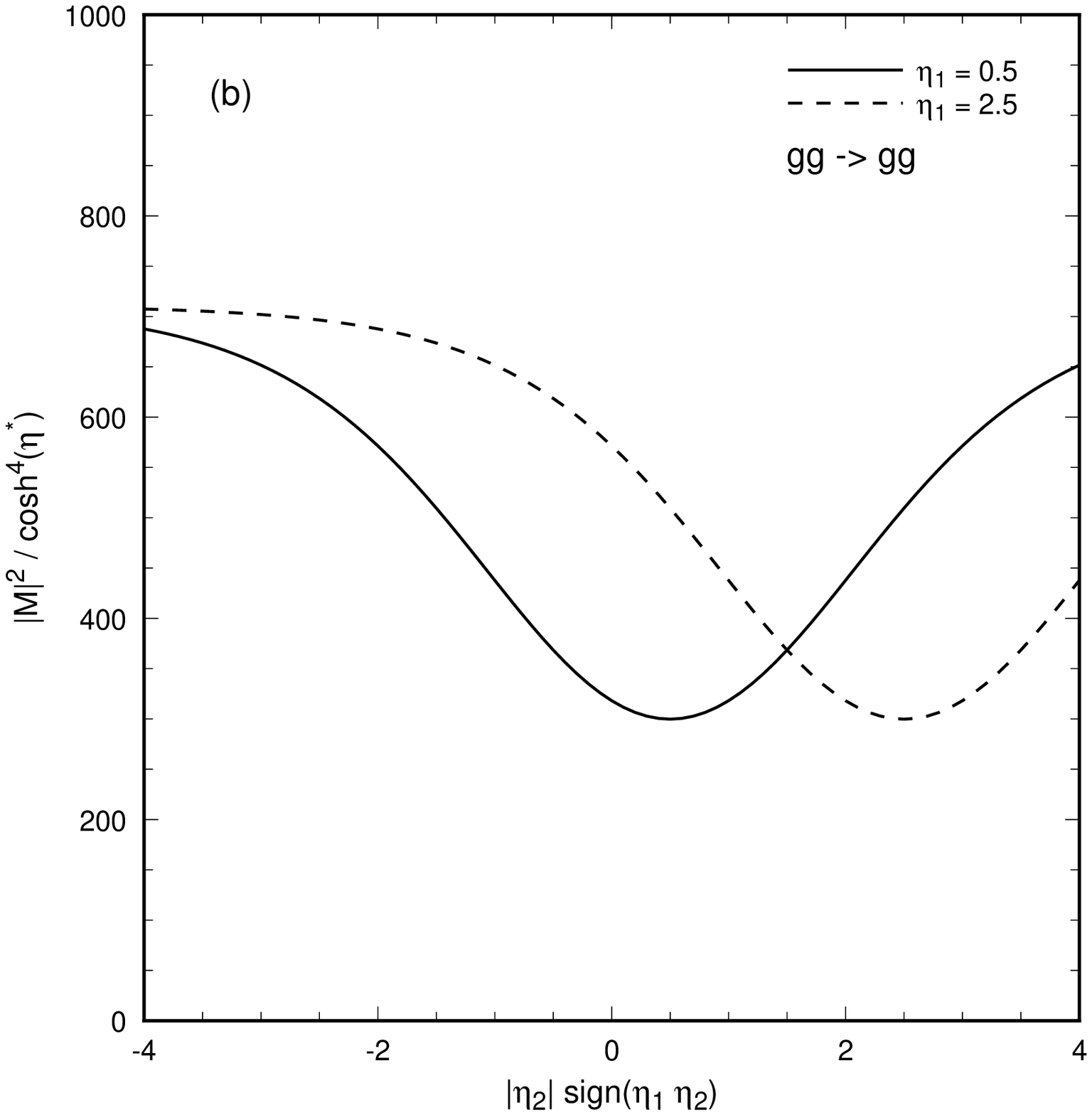}
\caption[]{The (a)  parton-parton luminosity for the \mrsdm\ parton densities
and (b) squared matrix elements in the `single effective subprocess
approximation' for $\eta_1 = 0.5$ and $E_T = 50$~GeV (solid) and
$\eta_1= 2.5$ and $E_T = 60$~GeV (dashed). The factorisation scale is
chosen to be 50 GeV.}
\end{figure}

As suggested by the solid line in Fig.~9, the next-to-leading order
corrections reduce the cross section uniformly by about $10\%$ until
the kinematic limit on $\eta_2$ from the lowest order process is
approached.  We also see that the difference between the
\mrsdz\ and \mrsdm\ predictions is
about $23\%$ at $|\eta_2| \sim 0$ as expected from Fig.~8.  

At larger pseudorapidities, the next-to-leading order predictions 
give a much better description of the data than at leading order. The
preliminary data appear to favour the \mrsdz\ parameterisation at the
$x$ values probed here, $x\sim 0.05$.  
However, the errors are still large, and as mentioned
at the end of section~3, there are significant
uncertainties in the overall normalisation of the experimental data.

As discussed in the previous sections, there is also an uncertainty 
in the normalisation of the theoretical cross section due to the 
choice of renormalisation and factorisation scales.
This is particularly evident for the signed distribution since 
the lowest order cross section is proportional to $\alpha^2_s(\mu_R)$.
Even at next-to-leading order, the overall normalisation is still uncertain.
However, one would expect that the shape of the distribution is relatively  
insensitive to varying $\mu_R$.
This is illustrated in Fig.~13, which
shows the ratio of next-to-leading order
 predictions for different scales relative to
the next-to-leading order \mrsdm\ prediction for $\mu = E_{T1}$.  
\begin{figure}
\vspace{8cm}
\includegraphics{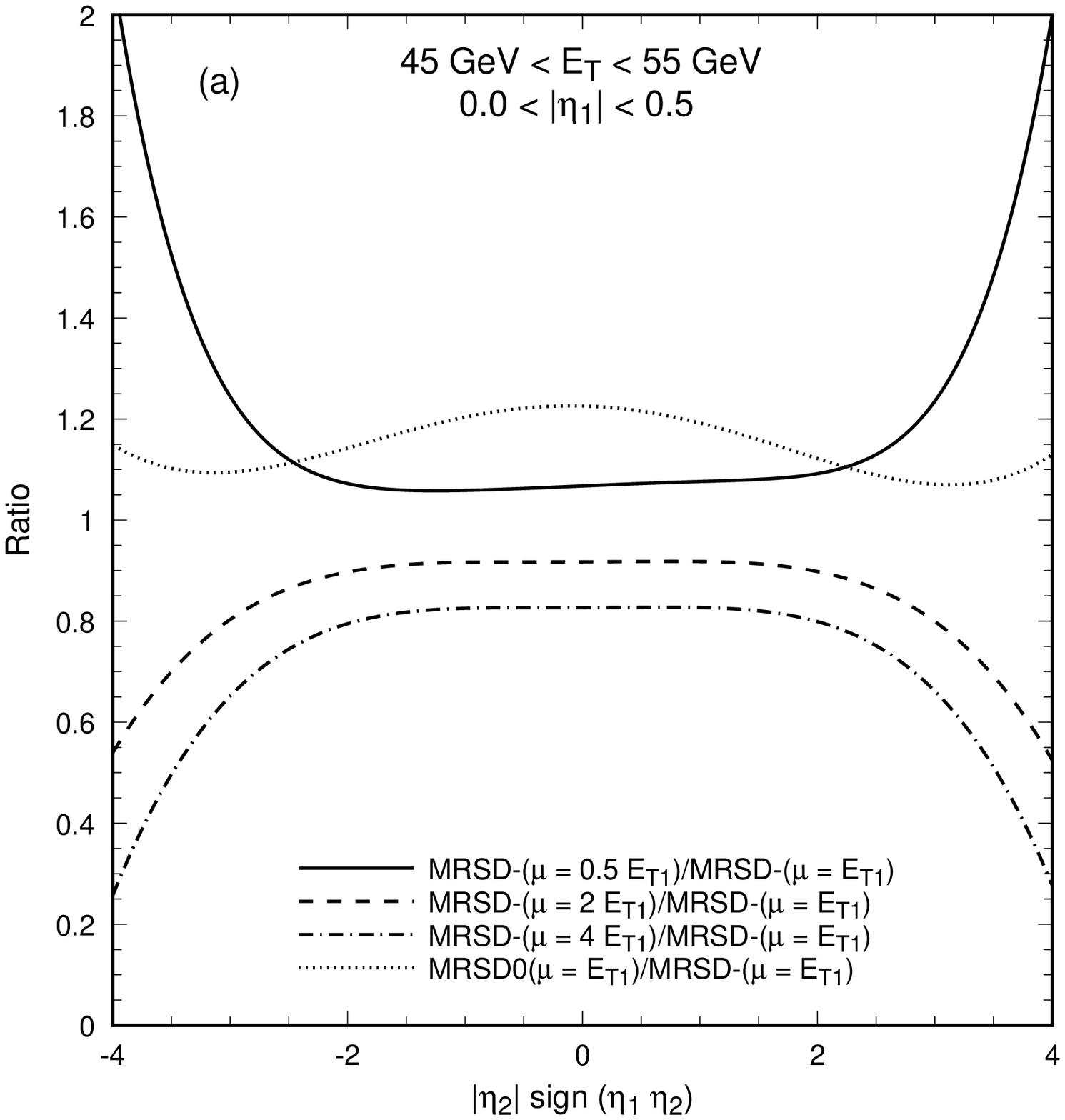}
\includegraphics{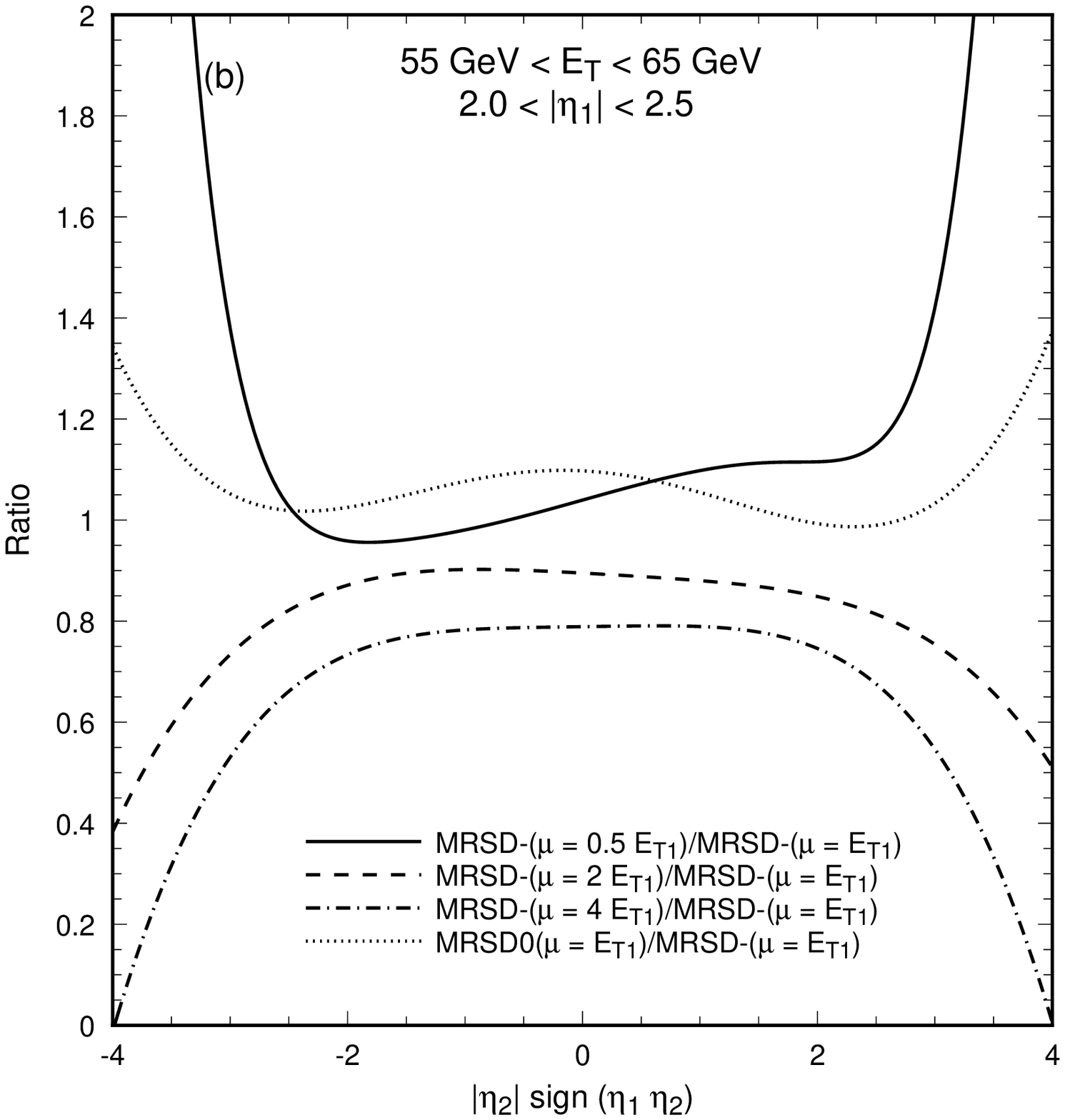}
\caption[]{The ratio of next-to-leading order predictions for the
signed distribution for the \mrsdm\ parton densities evaluated at $\mu =
\lambda E_{T1}$ for $\lambda=0.5$ (solid), $\lambda=2$ (dashed) and
$\lambda=4$ (dotdashed) relative to that for $\mu =E_{T1}$ for (a) 45~GeV
$ < E_T < 55$~GeV and $0.0 < |\eta_1| < 0.5$ and (b) 55~GeV $ < E_T <
65$~GeV and $2.0 < |\eta_1| < 2.5$.  The ratio of next-to-leading order
predictions for the \mrsdz\ and \mrsdm\ parton densities with
$\mu = E_{T1}$ is shown as a dotted line.}
\end{figure}

As expected, the
normalisation of the small $|\eta_2|$ region is quite sensitive to the
choice of scale, however, the {\em shape} of the distribution for small
$|\eta_2|$ is essentially unchanged for the central
$\eta_1$ slice, $0.0 < |\eta_1| < 0.5$.  For large pseudorapidities,
$|\eta_2| > 3$, there is of course a large scale dependence.  This is 
a consequence of exceeding 
the lowest-order kinematic limit on $\eta_2$; in this region, an
${\cal O}(\alpha_s^3)$ calculation such as the one performed here is
in fact a {\it leading\/}-order one.
As a contrast, the ratio of the next-to-leading order 
predictions for the \mrsdz\ and \mrsdm\ parton densities at
$\mu = E_{T1}$ are also shown.  In addition to a sizeable change in the
normalisation, the shape of the distribution around $|\eta_2| \sim 0$
is also changed.  It remains an experimental question as to whether
this difference in shape can be detected.

\section{The CDF same-side over opposite-side ratio}

The interpretation of D0 signed distribution measurement hinges
strongly on the absolute normalisation of the cross section. Thus one
would need to know the jet energy
correction well before one can constrain the parton density functions.
To circumvent this problem the CDF collaboration
has considered a ratio, that of
same-side (SS) to opposite-side (OS) cross sections~\cite{CDFssos}.  For the
same-side cross section, both jets have roughly the
same pseudorapidity, while in
the opposite-side cross section the jets are required to have roughly equal,
but opposite pseudorapidities.  One then forms this ratio, as a function
of the pseudorapidity in several tranverse-energy slices.
In a realistic experimental analysis, the pseudorapidities and
transverse energies will be binned so that,
\begin{eqnarray}
\sigma_{SS}(\eta)\Big \rfloor_{E_{T{\rm min}}<E_T<E_{T{\rm max}}} &=&
\int_{\eta-\Delta \eta}^{\eta+\Delta \eta} d\eta_1 
\int_{\eta-\Delta \eta}^{\eta+\Delta \eta} d\eta_2
\int_{E_{T{\rm min}}}^{E_{T{\rm max}}} dE_T 
\frac{d^3\sigma}{ dE_Td\eta_1 d\eta_2}, \\
\sigma_{OS}(\eta)\Big \rfloor_{E_{T{\rm min}}<E_T<E_{T{\rm max}}} &=&
\int_{\eta-\Delta \eta}^{\eta+\Delta \eta} d\eta_1 
\int_{-\eta-\Delta \eta}^{-\eta+\Delta \eta} d\eta_2
\int_{E_{T{\rm min}}}^{E_{T{\rm max}}} dE_T 
\frac{d^3\sigma}{ dE_Td\eta_1 d\eta_2 }.
\end{eqnarray}
{}From these cross sections we form the SS/OS ratio,
\begin{equation}
R_{SS/OS}(\eta)\Big \rfloor_{E_{T{\rm min}}<E_T<E_{T{\rm max}}}=\frac
{\sigma_{SS}(\eta)\Big \rfloor_{E_{T{\rm min}}<E_T<E_{T{\rm max}}}}
{\sigma_{OS}(\eta)\Big \rfloor_{E_{T{\rm min}}<E_T<E_{T{\rm max}}}},
\end{equation}
with the advantage that a large part of the experimental and
theoretical uncertainties cancel.  However, most of the dependence on
the parton densities in the central region where $\eta_1\sim \eta_2
\sim 0$ is also removed.  As we saw in the previous sections, this is
exactly the region where there can be a strong dependence on the
parton density functions.  Nevertheless, we can still study the
behaviour of the gluon density at small $x$ by examining the SS/OS
ratio at large pseudorapidity. Since the $x$ values probed are much smaller
than in the signed pseudorapidity distribution,
typically $x\sim 4E_T^2/s$ rather than $x \sim 2E_T/\sqrt{s}$,
studying this ratio is to a large extent complementary to studying the
D0 signed pseudorapidity distribution.

In the preliminary CDF measurement \cite{CDFssos}, the jet transverse
energy was chosen to lie in four separate bins, 27 GeV $<E_T<$ 60 GeV,
60 GeV $<E_T<$ 80 GeV, 80 GeV $<E_T<$ 110 GeV and 110 GeV $<E_T<$ 350
GeV and the pseudorapidity interval to be $\Delta\eta = 0.2$. It may prove
possible to extend the analysis to smaller transverse energy and
include a fifth bin, 15 GeV $< E_T < $ 27 GeV. 
As for the triply-differential and signed rapidity distributions,
we must symmetrize explicitly over the leading and next-to-leading transverse
energy jets in order to ensure that the sided cross sections
are infrared safe. In addition, in order to suppress
events with three or more hard jets, an azimuthal angle cut between
the two leading jets of $\pi-0.7<\Delta\phi<\pi+0.7$ is applied.

\begin{figure}[t]\vspace{8cm}
\includegraphics{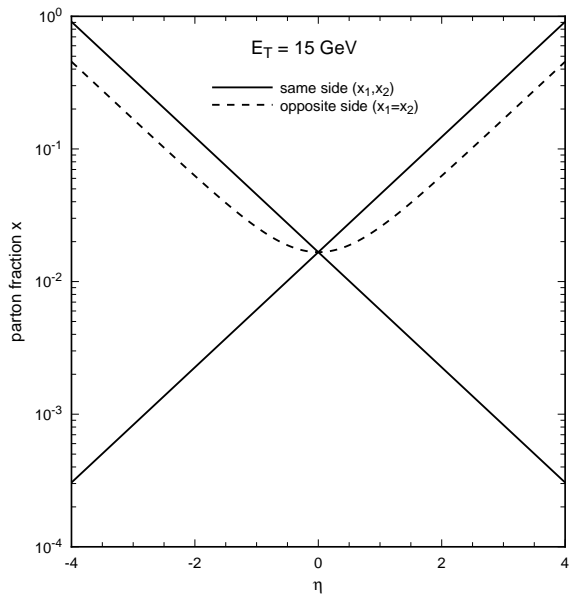}
\caption[]{The leading order parton fractions probed by the SS/OS 
cross section ratio for $E_T = 15$~GeV.}
\end{figure}

\begin{figure}[t]\vspace{8cm}
\includegraphics{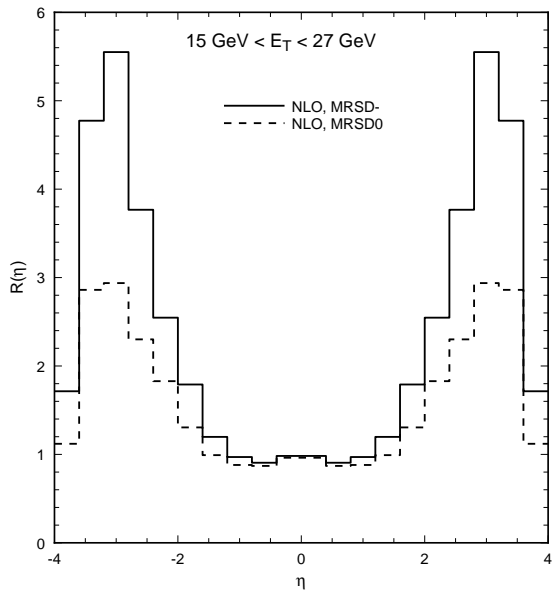}
\caption[]{
The next-to-leading order predictions for the SS/OS ratio evaluated at
$\mu = E_{T1}$ for the \mrsdm\ (solid), \mrsdz\ (dotted) and MRSA (dashed)
parton distributions as a function of $\eta$ for the smallest $E_T$
bin accessible to the CDF collaboration, 15~GeV $< E_T < 27$~GeV.}
\end{figure}

To get a feeling for the range of parton fractions probed by this
particular cross section, Fig. 14 shows the three different momentum
fractions (using the leading order definition of Eq. 1) as a function
of the pseudorapidity for the smallest accessible jet transverse energy $E_T
= 15$~GeV.  
We see that it is possible to probe parton fractions 
for $x$ values as small as $3\times 10^{-4}<x<1$. 

The next-to-leading order predictions for the SS/OS ratio
for the very low transverse energy bin, 15~GeV $< E_T < 27$~GeV, are
shown in Fig.~15.  Because very low $x$ values are encountered, the
different parton density functions give a broad range of predictions.
At present, no data are available for this particular transverse energy
range. However, even with relatively large experimental uncertainties
one can still easily discriminate between different parton densities.
This makes the very small $E_T$ bin particularly interesting. 

The region around $\eta=3$ has the largest
sensitivity to different parton density functions, corresponding to a
smallest parton fraction of $8\times 10^{-4}$.  At larger pseudorapidities
(and therefore smaller $x$ values) there is a severe phase space
supression and the cross section (and event rate) decreases rapidly.

\begin{figure}[t]\vspace{12cm}
\includegraphics{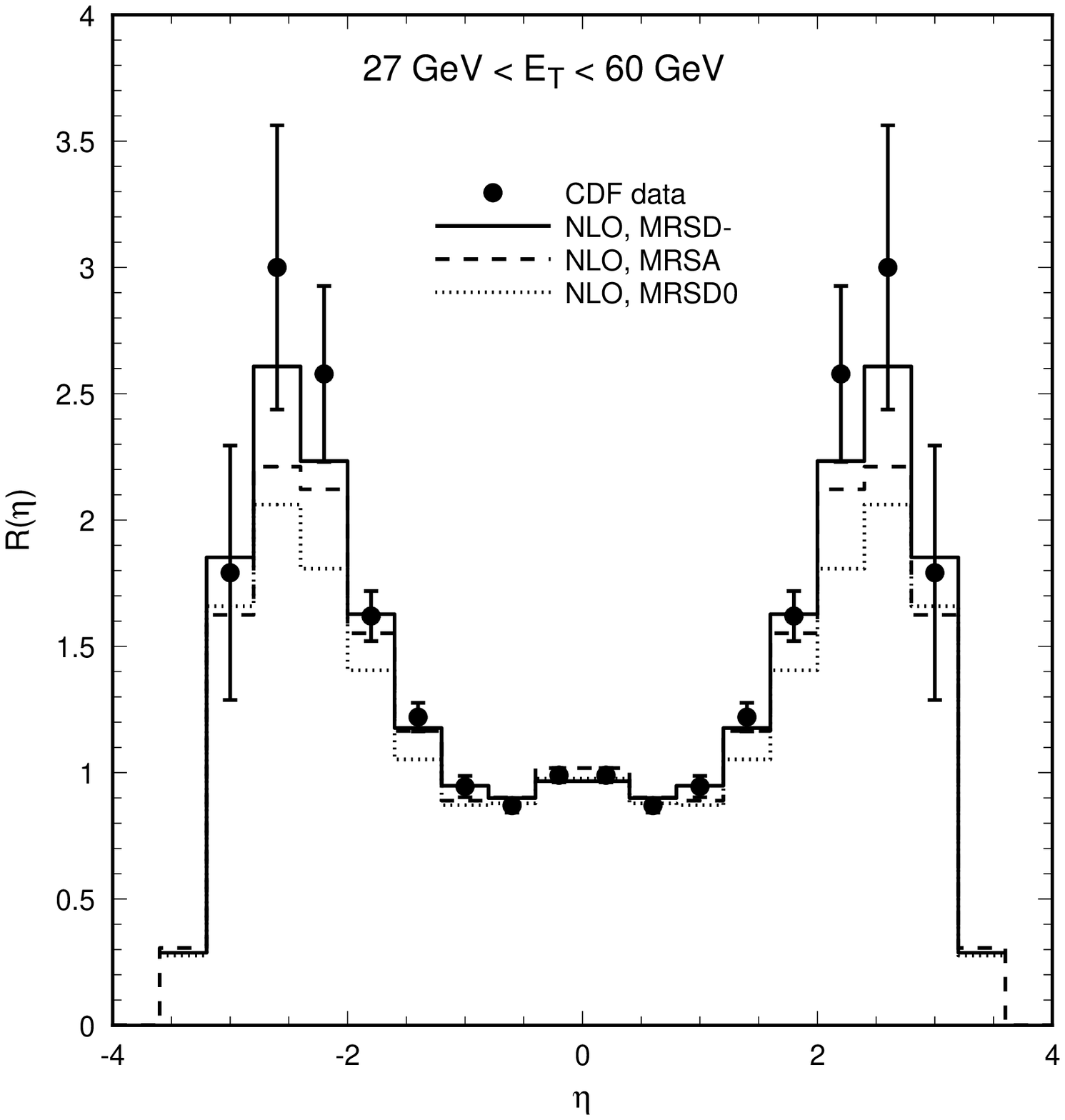}
\includegraphics{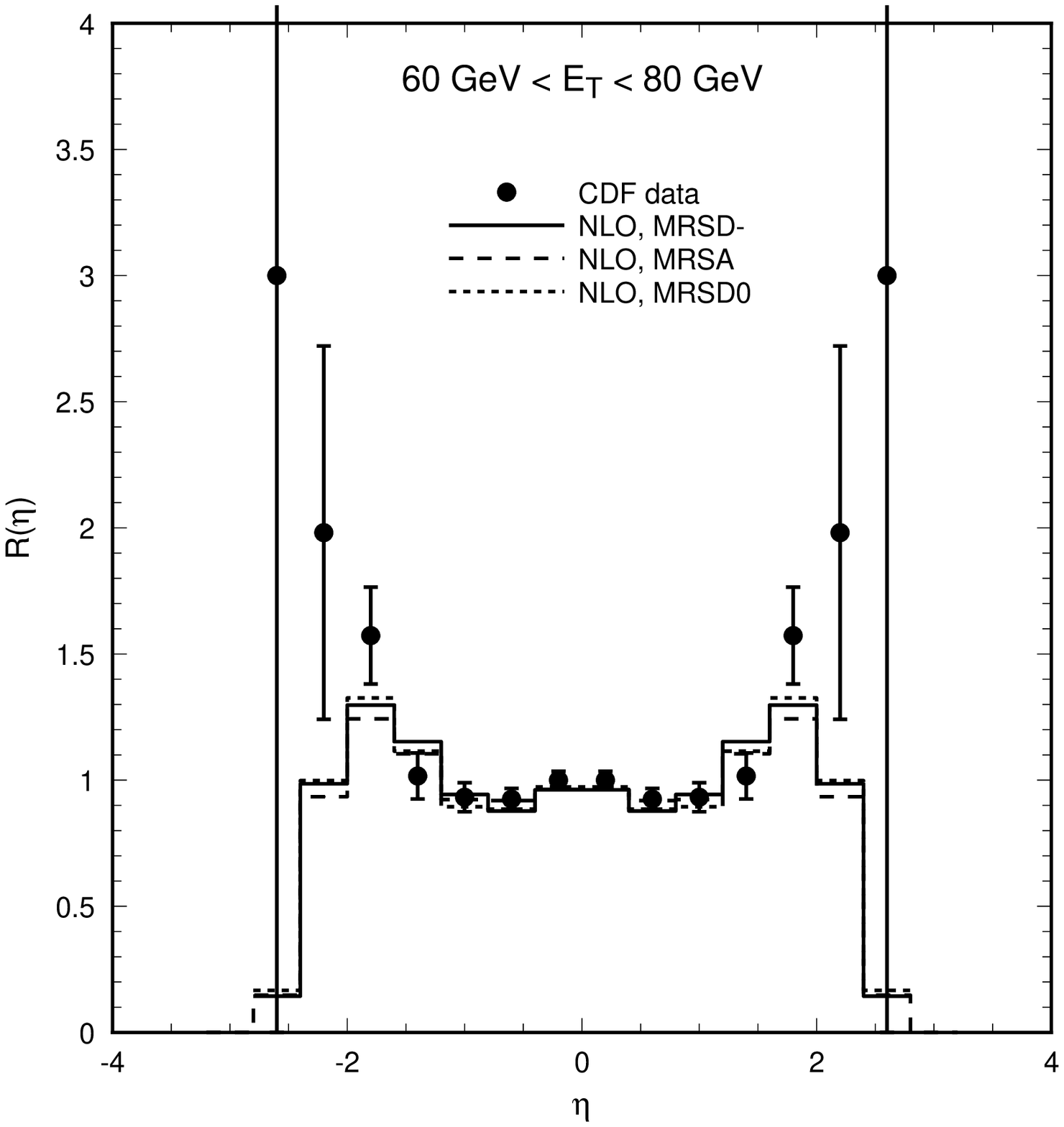}
\includegraphics{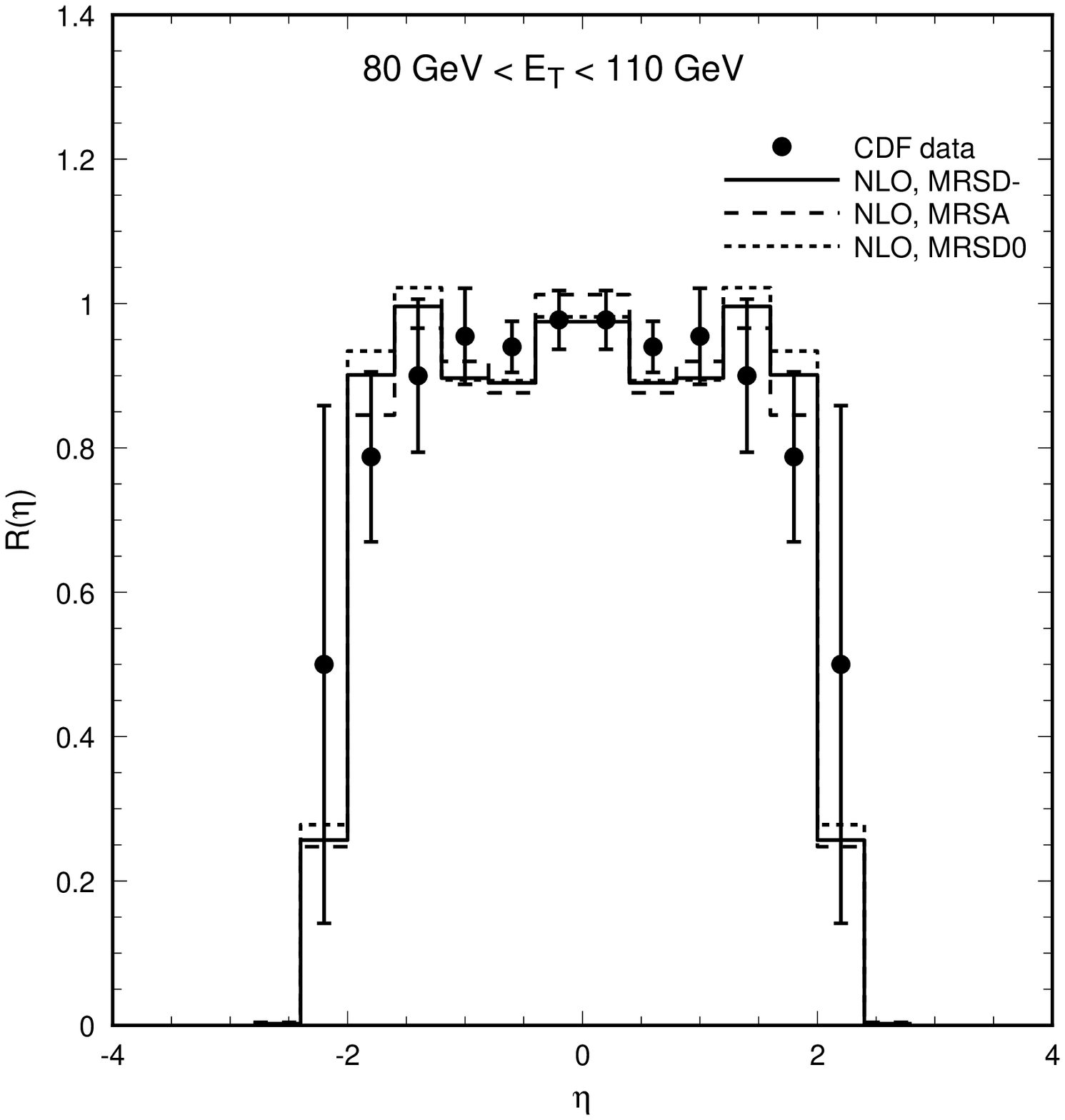}
\includegraphics{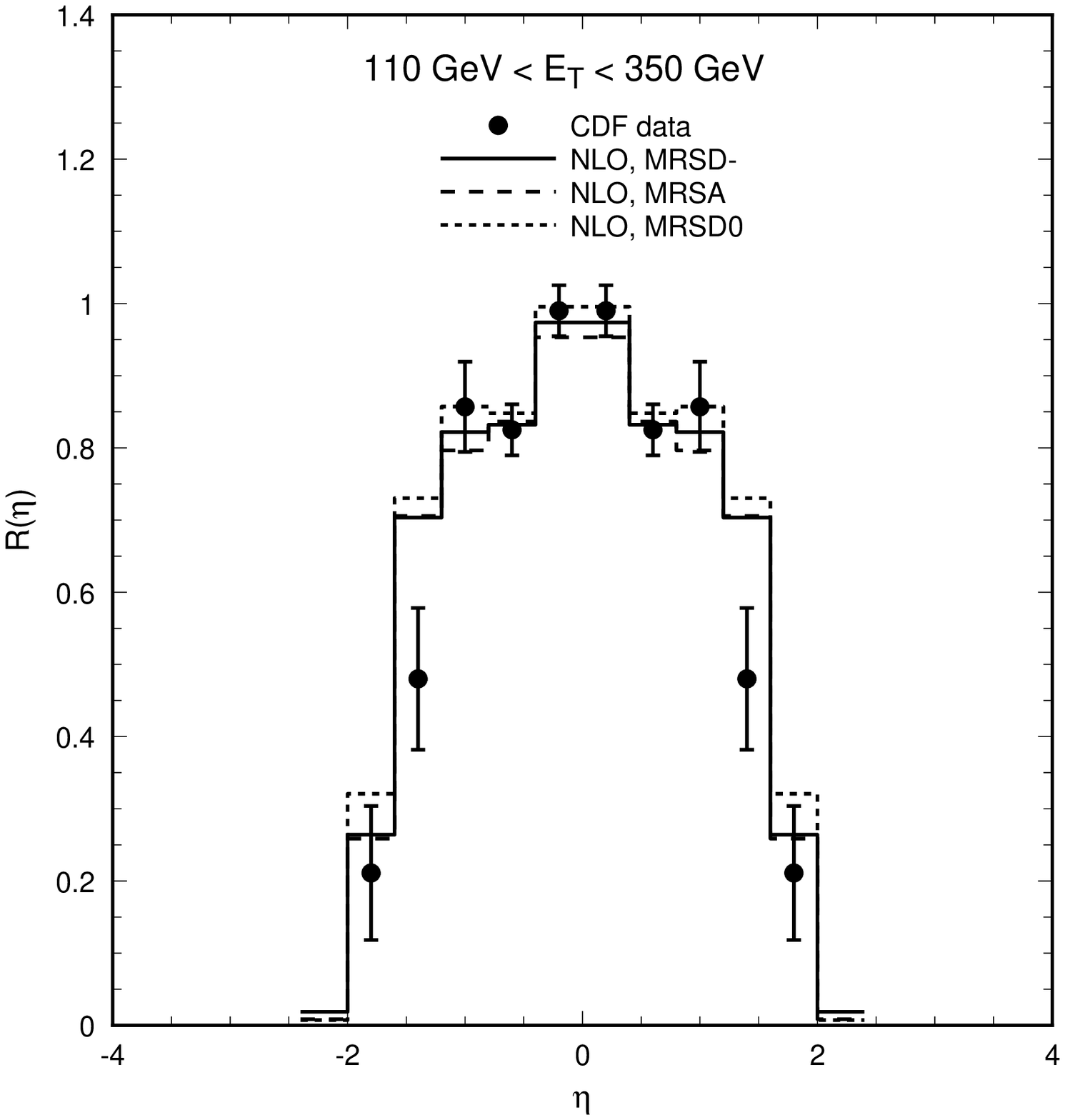}
\caption[]{The next-to-leading order (NLO) predictions for the SS/OS ratio
evaluated at $\mu = E_{T1}$ for the \mrsdm\ (solid), \mrsdz\ (dotted) and
MRSA (dashed) parton distributions as a function of $\eta$ for the
four transverse energy bins, 27 GeV $<E_T<$ 60 GeV, 60 GeV $<E_T<$ 80
GeV, 80 GeV $<E_T<$ 110 GeV and 110 GeV $<E_T<$ 350 GeV.  Also shown
is the preliminary CDF data of ref. \cite{CDFssos}}
\end{figure}

In the other four transverse energy bins, the CDF collaboration
has published preliminary
data which we display along with next-to-leading order predictions for
three different parton density sets in Fig.~16.  Because the
experimental results are preliminary, we should be careful in drawing
conclusions from these results for the time being.  
These results do however prefigure discriminatory powers which should
emerge as the experimental errors shrink with the inclusion of more
data from the current Tevatron run.
Even with the current uncertainties, the
smallest $E_T$ bin suggests a deficiency in the
density of \mrsdz\ partons round $x \sim 10^{-3}$, and in contrast 
to the signed rapidity distribution discussed in section~4, favors
the \mrsdm\ set of structure functions.  

For the higher
transverse energy bins higher $x$ values are sampled and there is not
much difference between the different parton density functions.  
As an example, the highest
transverse energy bin is mostly sensitive to parton momentum fractions around
$0.1$, where the parton density functions are, in principle, tightly
constrained.  Of course, one should still compare the data with
theory. That different parton densities give the same results
does not guarantee they are correct; it merely indicates that they either use
the same data as a constraint or use the same assumptions to derive
the individual parton densities.

\begin{figure}[t]\vspace{8cm}
\includegraphics{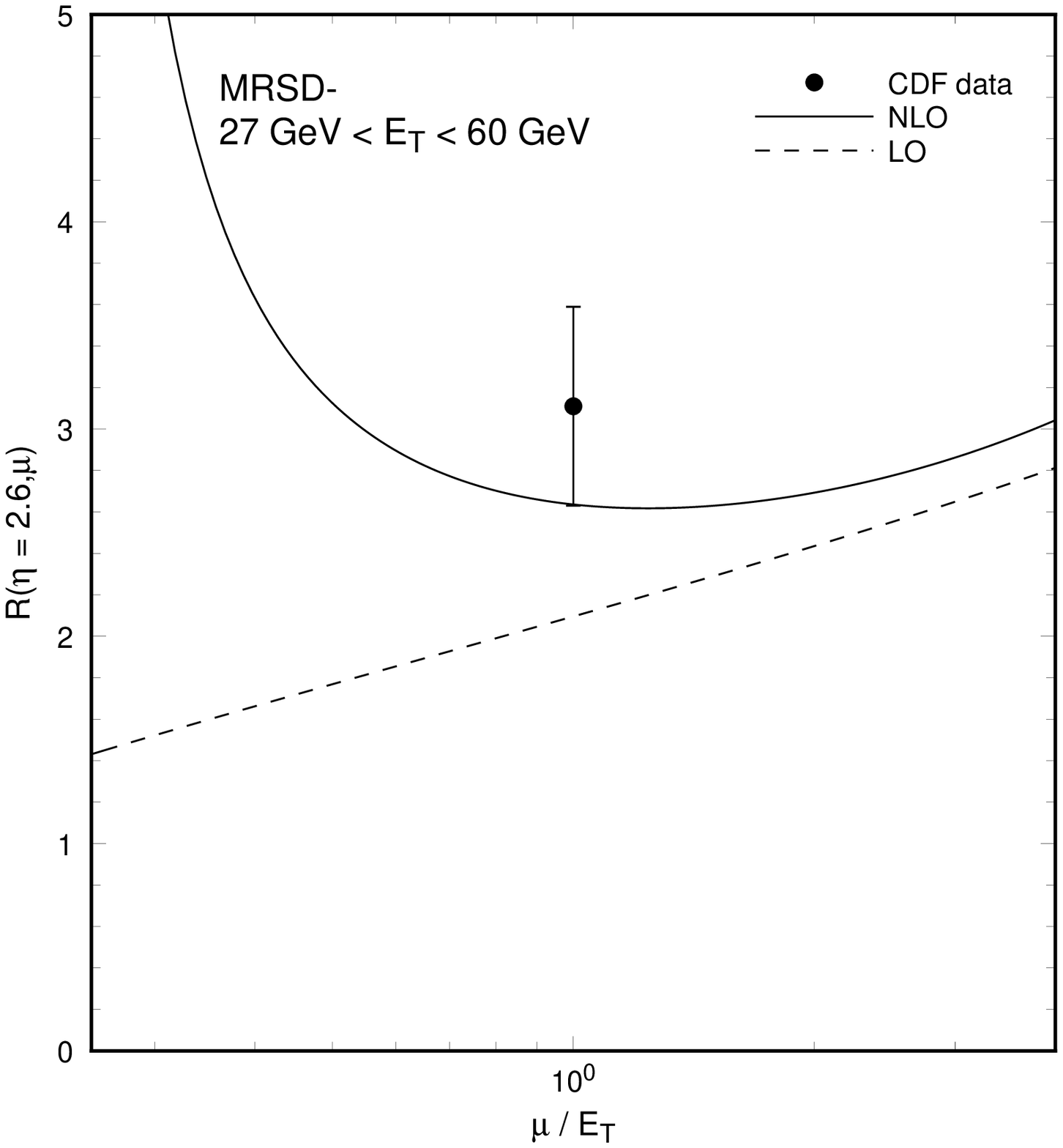}
\caption[]{The renormalisation/factorisation scale dependence 
of the SS/OS ratio for $\eta=2.6$ in the 27 GeV $< E_T <$ 60 GeV bin
using the \mrsdm\ structure functions.  The data point is taken from
ref.~\cite{CDFssos}.}
\end{figure}

For this distribution as well, we should also consider the uncertainty
 in the theoretical predictions arising from 
renormalisation and factorisation scale dependence present in perturbative
QCD calculations.  To study this, Fig.~17 shows the SS/OS ratio for
the 27 GeV $< E_T <$ 60 GeV transverse energy bin at pseudorapidity
$\eta=2.6$ as a function of $\mu = \mu_R = \mu_F$.  This phase space
point offers the highest discriminatory power with the currently
available data.  As an illustration we have chosen the \mrsdm\ set which
seems to be favoured by these data.  The reference scale is the
transverse energy of the highest $E_T$ jet in the event, $\mu =
E_{T1}$, as in the previous sections.  The leading-order prediction
does not depend on the strong coupling constant.  As the SS and OS
cross sections probe the parton densities at different momentum fractions~$x$,
it does depend on the factorization scale, but this gives rise only to
a trivial scale dependence: the SS to OS ratio rises linearly with
$\log(\mu/E_{T1})$.  At next-to-leading order, the prediction does depend
on the coupling constant, and the overall dependence is less trivial.
It so happens that the reference scale, $\mu = E_{T1}$, coincides with
the minimum in the variation of the ratio with respect to scale.  
Either increasing or decreasing the scale results in
an increase of the SS/OS ratio. Varying the scale by a factor of
two around the reference scale, we get a feeling for the
theoretical uncertainty. The largest variation comes from reducing $\mu$ and
increases the cross section by approximately 20\%. Due to this
uncertainty one cannot really discriminate between the \mrsdm\
distributions with $xg(x) \sim x^{-0.5}$ and the more recent MRSA fits
with $xg(x) \sim x^{-0.3}$.  However, it should be possible to exclude
the \mrsdz\ distributions once the final CDF data are published. To
discriminate between \mrsdm\ and MRSA, it will be necessary to make a
measurement in the 15 GeV $< E_T <$ 27 GeV transverse energy bin.

\begin{figure}[t]\vspace{8cm}
\includegraphics{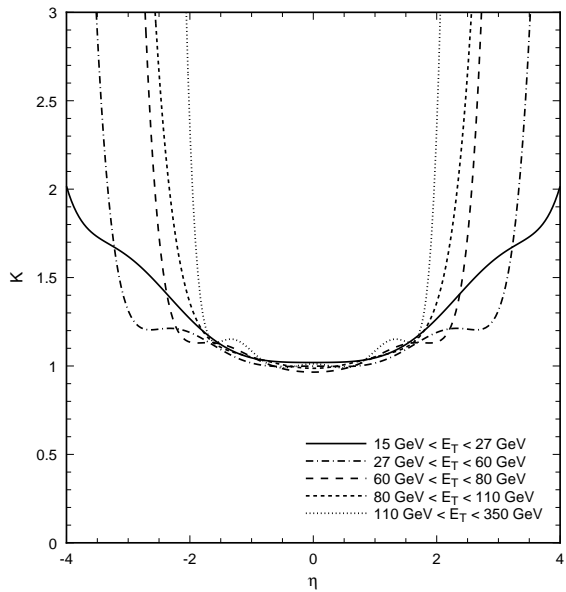}
\caption[]{The ratio of next-to-leading to leading order predictions
for the SS/OS ratio with $\mu=E_{T1}$ and \mrsdm\ parton densities in
the five transverse energy intervals.}
\end{figure}

Fig.~18 shows the ratio of next-to-leading to leading order
predictions (the $K$-factor) for 27 GeV $< E_T <$ 60 GeV and 80 GeV $<
E_T <$ 110 GeV using the \mrsdm\ parton densities and $\mu = E_{T1}$.
The shape of the SS/OS ratio is basically unchanged between
$-2\leq\eta\leq 2$. For larger pseudorapidities we get a rapid change in the
$K$-factor. This is again mainly due to the fact that the leading order
cross section is quickly forced to zero by the kinematic constraints on 
$2 \rightarrow 2$ scattering, and does {\em not\/} indicate the
presence of large logarithms which might spoil the applicability
of perturbation theory.  We have parameterised the
$K$-factor as an even polynomial in $\eta$,
\begin{equation}
K(\eta)=A+B\eta^2+C\eta^4+D\eta^6+E\eta^8.
\end{equation}
The fitted constants $A\ldots E$ for all five transverse energy
intervals are given in Table~1.  We see that for this double ratio,
the corrections are extremely small in the central region.

\begin{table}\begin{center}
\begin{tabular}{|c|lllll|} \hline
Pseudorapidity bin &\multicolumn{1}{c}{A}&  
\multicolumn{1}{c}{B}&\multicolumn{1}{c}{C}
&\multicolumn{1}{c}{D}&\multicolumn{1}{c|}{E}\\ \hline
15 GeV $< E_T <$ 27 GeV   & 1.0197 &  0.0062 &  0.0226 & -0.0025 & 8.166e-5 \\
27 GeV $< E_T <$ 60 GeV   & 0.9992 & -0.019  &  0.0453 & -0.0094 & 5.592e-4 \\
60 GeV $< E_T <$ 80 GeV   & 0.9653 &  0.072  &  0.0326 & -0.0208 & 0.0027   \\
80 GeV $< E_T <$ 110 GeV  & 0.9869 &  0.0891 & -0.0332 &  0.0057 & 6.427e-4 \\
110 GeV $< E_T <$ 350 GeV & 1.0121 & -0.1974 &  0.4985 & -0.2743 & 0.0457   \\
\hline
\end{tabular} \end{center}
\caption[]{The parameterisation of the $K$-factor for \mrsdm\ 
in the 5 transverse energy bins. The fitted formula is
$K(\eta)=A+B\eta^2+C\eta^4+D\eta^6+E\eta^8$ and is intended to be used
for returning the $K$-factor for the bin, given the center of the bin
(that is $\eta$ = 0.2, 0.6, 1.0, $\ldots$)}
\end{table}

\newpage

\section{Conclusions}

In this paper we have made a detailed study of the two-jet cross
section in hadron-hadron collisions for a given range of jet
transverse energy as a function of the pseudorapidities of the two
jets.  This distribution is particularly sensitive to the parton
density functions at small ($x\sim {\rm few}~\times~10^{-3}$) to intermediate
($x\sim {\rm few}~\times~10^{-2})$ parton momentum fractions for jet
transverse energies accessible to the CDF and D0 experiments at
Fermilab.  For example, parton distributions that are relatively large
at small $x$ (and therefore constrained to be relatively small at
larger $x$ by the momentum sum rule) lead to a relative enhancement of
the same-side two-jet cross section at large pseudorapidities ($\eta_1
\sim \eta_2 \gg 0$) and a relative depletion for central jets, $\eta_1
\sim \eta_2 \sim 0$.  We have studied the next-to-leading QCD
corrections to this distribution in some detail.  To discuss the
properties of the cross section it is convenient to divide the $\eta_1
- \eta_2$ plane into two regions, the central region where the
pseudorapidity of the two jets is less than the lowest order kinematic
limit, $|\eta_1|,~|\eta_2| < \cosh^{-1}(1/x_T)$ and the foward region
where the rapidity of one of the jets approaches or exceeds the lowest
order boundary.  In the central region,
\begin{description}
\vspace{-0.3cm}
\item{i)} the next-to-leading order corrections are small 
and perturbation theory works well.
\item{ii)} the scale uncertainty in the overall normalisation 
is reduced.  Varying $\mu$ by a factor of 2 about $\mu=E_{T1}$ 
changes the next-to-leading order prediction by ${\cal O}(8\%)$.
\item{iii)} the scale uncertainty in the shape of the distribution 
is quite small.
For the same variation of $\mu$, the relative bin-to-bin correction 
is less than ${\cal O}(2\%)$ --- see Fig.~10.
\end{description}
On the other hand, in the forward region,
\vspace{-0.3cm}
\begin{description}
\item{i)} the next-to-leading order corrections are important 
and improve the agreement with experimental data.
\item{ii)} there is a considerable scale uncertainty because 
the corrections calculated in a `next-to-leading' order program
corrections are effectively lowest order.
\end{description}
Because the central pseudorapidity region is both sensitive to the
parton densities and stable to higher order corrections, the triply
differential distribution offers an excellent chance to gain extra
information about the distribution of partons in the proton.  The
current and future runs of the Tevatron at Fermilab should yield
copious quantities of two-jet events and once precise experimental
data are available, it should be possible to make a determination of
the gluon density at small and intermediate $x$ values.  However, in
addition to the uncertainty in the normalisation of the theoretical
predictions, there is also an uncertainty in the experimental
normalisation due to uncertainties in the luminosity measurement, 
in the jet energy calibration, and other effects.  
The information on the parton densities lies more in
the {\em shape\/} than the overall normalisation, and one way of determining
the parton densities is to allow the overall normalisation of the
theoretical prediction, $\sigma^{TH}$, to float, so that by varying
$c$ the $\chi^2$ for
$$
\int^{E_{T{\rm max}}}_{E_{T{\rm min}}}  dE_T
\left (\frac{d^{3}\sigma^{EXP}}{dE_Td\eta_1d\eta_2} -
c \frac{d^{3}\sigma^{TH}}{dE_Td\eta_1d\eta_2}\right),
$$
summed over the different $\eta_1$, $\eta_2$ cells in the central
pseudorapidity region is minimised for a given parameterisation of the
parton densities.  By adjusting the input parameterisations, the
$\chi^2$ may be further reduced as the predicted shape becomes closer
and closer to that observed in the data.  Of course, data can be taken
for many slices in transverse energy.  This allows the possibility of
following the evolution of the parton densities directly by following
trajectories of constant $(x_1,x_2)$ in the $\eta_1-\eta_2$ plane as a
function of $E_T$.

At present only preliminary data are available for particular slices of
the $\eta_1-\eta_2$ plane.  The signed distribution (sect.~4)
presented by the D0 collaboration is primarily sensitive to
intermediate $x$ values.  As we may expect from the preceding discussion,
the shape is well predicted, but the normalisation is uncertain.
Nevertheless, the preliminary data appear to favour the \mrsdz\
parameterisations (Fig.~11).  On the other hand, the same-side to
opposite-side cross section ratio presented by the CDF collaboration
probes much smaller $x$ values.  Once again, the shape is relatively
unchanged by including the next-to-leading order corrections.  As
shown in Fig.~16, the preliminary data (again with large errors)
appear to favour the more singular \mrsdm\ parameterisation.  Of
course, with the current experimental data sample, no definitive
conclusion can be drawn.  However, if these tentative observations are
accurate, it would imply that the data favour the parton density with the 
largest density of gluons at {\em both} $x\sim {\rm few}~\times 10^{-3}$ 
{\em and} $x\sim {\rm few}~\times 10^{-2}$.
In other words, there are more gluons present in the
small and intermediate $x$ regions than expected from the momentum sum
rule, which may in turn suggest that the density of gluons is not as
well determined by direct photon data (WA70) as previously thought.  Data
from the current Tevatron run should help to provide an answer to this puzzle.

\section*{Acknowledgements}

We are happy to acknowledge many stimulating discussions with members
of the D0 and CDF collaborations, in particular, Brad Abbott, Jerry
Blazey, Teri Geld, Eve Kovacs and Freedy Nang.  We also thank Alan
Martin for constructive comments.  EWNG thanks the Fermilab theory
group for its kind hospitality in July and August when this work was
initiated.

\end{document}